\documentclass[usenatbib]{mnras}
\usepackage{epsf,graphics,graphicx}
\usepackage{amsmath}
\usepackage{amssymb,latexsym,mathrsfs, bm}
\usepackage{color}
\usepackage{float}
\usepackage{natbib}
\usepackage{times}
\usepackage{pdflscape}
\usepackage{longtable}

\def\eq#1{{Eq.~(\ref{#1})}}

\title[Intensity mapping forecasts]{Impact of astrophysics on cosmology forecasts for 21 cm surveys}

\author[Padmanabhan, Refregier and Amara]{Hamsa
Padmanabhan$^{1,2}$\thanks{Electronic address: hamsa@cita.utoronto.ca},
Alexandre Refregier$^1$\thanks{Electronic address:
{alexandre.refregier@phys.ethz.ch}},
Adam Amara$^1$\thanks{Electronic address: {adam.amara@phys.ethz.ch}
}\\
$^{1}$ Institute for Particle Physics and Astrophysics, ETH Zurich, Wolfgang-Pauli-Strasse 27, CH-8093 Z\"{u}rich, Switzerland\\
$^{2}$ Canadian Institute for Theoretical Astrophysics, 60 St. George St., Toronto, ON, M5S 3H8, Canada}

\begin{document}
\date{ }
\maketitle

\begin{abstract}
We use the results of previous work building a halo model formalism for the  distribution of neutral hydrogen, along with experimental parameters of future radio facilities, to place forecasts on  astrophysical and cosmological parameters from next generation surveys. We consider 21 cm intensity mapping surveys conducted using the  BINGO, CHIME, FAST, TianLai, MeerKAT and SKA experimental configurations. We work with the 5-parameter cosmological dataset of \{$\Omega_m, \sigma_8, h, n_s, \Omega_b$\} assuming a flat $\Lambda$CDM model, and the astrophysical parameters \{$v_{c,0}, \beta$\} which represent the cutoff and slope of the HI- halo mass relation. 
We explore (i) quantifying the effects of the astrophysics on the recovery of the cosmological parameters, (ii) the dependence of the cosmological forecasts on the details of the astrophysical parametrization, and (iii) the improvement of the constraints on probing smaller scales in the HI power spectrum.  For an SKA I MID intensity mapping survey alone, probing scales up to $\ell_{\rm max} = 1000$, we find a factor of $1.1 - 1.3$ broadening in the constraints on  $\Omega_b$ and $\Omega_m$, and of $2.4 - 2.6$ on $h$, $n_s$ and $\sigma_8$,  if we marginalize over astrophysical parameters without any priors. However, even the prior information coming from the present knowledge of the astrophysics largely alleviates this broadening. These findings do not change significantly on considering an extended HIHM relation, illustrating the robustness of the results to the choice of the astrophysical parametrization. Probing scales up to $\ell_{\rm max} = 2000$ improves the constraints by factors of 1.5-1.8. The forecasts improve on increasing the number of tomographic redshift bins, saturating, in many cases, with 4 - 5 redshift bins. We also forecast constraints for  intensity mapping with other experiments, and draw similar conclusions.
\end{abstract}

\begin{keywords}
cosmology:observations -- radio lines:galaxies -- cosmology:theory
\end{keywords}

\section{Introduction}

Upcoming and future radio experiments aim to probe the distribution of neutral hydrogen with its redshifted 21-cm line, both during the dark ages and cosmic dawn \citep[see, e.g.,][]{madau1997} as well as in the post-reionization universe \citep[e.g.,][]{chang10, masui13, switzer13}. In the latter case, recent work aims to use the intensity mapping technique \citep[e.g.,][]{bharadwaj2001, loeb2008}, for which the resolution of individual objects is not required and the power spectrum of the intensity fluctuations is directly measured. Many of the 21 cm experiments aim to measure fundamental physics parameters by placing constraints on, e.g., dark energy  \citep[e.g.,][]{bull2014}, modified gravity \citep[e.g.,][]{hall2013} or inflationary models \citep[e.g.,][]{xu2016}. In order to have a realistic estimate of the degree of cosmological information that can be extracted from  these experiments, it is important to quantify the extent of astrophysical degradation in these studies. This is an important effect which can be called the `astrophysical systematic',  and  has consequences for our  derivation of cosmological forecasts from the knowledge of the HI power spectrum.  

In \citet[][hereafter Paper I]{hptrcar2015}, we provided a quantitative estimate of the degree of this uncertainty, by using a minimum-variance estimator applied to the key astrophysical quantities that influence the HI power spectrum. We found that the astrophysical uncertainties cause the order of $60 - 100 \%$ uncertainty in the measured power spectrum. This can be further expressed as a function of redshift and the resulting estimates are provided in Table 3 of that paper. 

In follow-up analytical work to the theoretical and observational uncertainties above, we developed a halo model framework to understand the distribution and evolution of HI in the post-reionization universe, by considering the current data both from 21 cm intensity mapping and resolved emission, as well as from Damped Lyman-Alpha (DLA) systems \citep[][hereafter Paper II]{hparaa2016}. The parameters of this halo model were astrophysical, and related to how HI populates haloes both in terms of the HI mass - halo mass relation, as well as the HI radial distribution profile. {{The parameters used were: (i) the concentration normalization parameter, $c_{\rm HI, 0}$, (ii) the evolution of the concentration with redshift, specified by $\gamma$, (iii) the overall normalization for the $M_{\rm HI} - M$ relation, $\alpha$,  (iv) the slope of the relation, $\beta,$ and a lower cutoff in virial velocity, $v_{\rm c,0}$.}} Constraints on these astrophysical parameters were possible using the combined set of the low- and high redshift observations, and the statistical uncertainties resulted in fairly tight error bars on the estimation of the free parameters.

In this paper, we combine the understanding of the uncertainties in the astrophysics as described in Paper I, with the modelling framework for these as presented in Paper II, towards building realistic forecasts for current and future 21 cm experiments. In this work, we concentrate on the intensity mapping observations, in which the individual systems are not resolved. 
{{Typically, three-dimensional analyses of clustering in wide-field surveys require the assumption of an underlying cosmological model. This requirement is circumvented by performing a tomographic analysis with the angular correlation function (or power spectrum) within bins of redshift \citep[e.g.,][]{seehars2016, nicola2014}. Using the angular power spectrum, denoted by $C_{\ell} (z)$, is thus effectively suited to obtaining meaningful cosmological forecasts from an intensity mapping survey. This also facilitates the ease of cross-correlations between different probes \citep[e.g.,][]{eriksen2015}. We, therefore, use the angular correlation function as the primary measure of clustering from intensity mapping surveys in the present work.
}}

We first work with a `fiducial' configuration, taken to be the SKA I MID (using bands B1 and B2) and explore both (i) how the astrophysical uncertainties cause a degradation (`systematic') in the cosmological forecasts, and the extent to which this can be alleviated through tomography or the combining of redshift bins, and (ii) the constraints on the astrophysical parameters themselves, achievable with an intensity mapping survey. In this work, we consider the cosmology to be given by the flat $\Lambda$CDM model with the free parameters \{$\Omega_m, \sigma_8, h, n_s, \Omega_b$\}. We use the astrophysical parameters $\beta$ and  $v_{\rm c,0}$ for describing the HI-halo mass relation.

We next investigate the impact of extending the multipole range from $\ell_{\rm max} = 1000$ to $\ell_{\rm max} = 2000$, thereby probing more non-linear scales for the fiducial configuration. We then investigate the effects of an extended parametrization of the HI-halo mass relation, beyond that favoured by the current data, on the forecasts obtained.
We also explore the cases of other upcoming HI intensity mapping experiments, namely the CHIME, BINGO, TianLai, MeerKAT and FAST configurations.  We discuss how the astrophysical effects influence the recovery of the cosmological parameters in each case. We summarize our conclusions and discuss future prospects in the final section.

\section{Fisher matrix forecasts}
\label{sec:fisherforecasts}
Here, we present the formalism for forecasting the constraints on astrophysics and cosmology with the Fisher matrix. 

{{The halo model formalism \citep[e.g.,][]{seljak2000, peacock2000, cooray2002} has led to a successful description of dark matter properties by using the halo mass function and density profile to describe dark matter abundances and clustering.}}
Extending the halo model framework to describe HI, developed in, e.g., \citet{hpar2017,hparaa2016}, the average HI mass associated with a dark matter halo of mass $M$ at redshift $z$  is given by:
\begin{eqnarray}
M_{\rm HI} (M,z) &=& \alpha f_{H,c} M \left(\frac{M}{10^{11} h^{-1} M_{\odot}}\right)^{\beta} \nonumber \\
&& \exp\left[-\left(\frac{v_{c0}}{v_c(M,z)}\right)^3\right] \nonumber \\
\label{hihm}
\end{eqnarray}
where the three free parameters are (i) $\alpha$, the overall normalization factor, (ii) $\beta$, the slope of the HI - halo mass relation, and (iii) $v_{\rm c,0}$, which is a lower virial velocity cutoff for the dark matter halo of mass $M$ to be able to host HI. {{The quantity $f_{\rm H,c}$ denotes the cosmic hydrogen fraction, defined through $f_{\rm H,c} = \Omega_b (1 - Y_p)/\Omega_m$ where $Y_p = 0.24$ is the helium abundance.}} \footnote{This is the best-fitting HI-halo mass relation favoured by present-day experiments. {{The form of the HI-halo mass relation is also supported by the results of several simulations \citep[e.g.,][]{navarro2014} which find that almost all the HI in the post-reionization universe resides within dark matter haloes. }} For completeness, we also validate the robustness of our results to the choice of this parametrization by extending the functional form in Sec. \ref{sec:evolution}.}

The distribution of HI in the dark matter halo is described by a radial profile function, of the form:
\begin{equation}
\rho(r,M) = \rho_0 \exp(-r/r_s)
\label{rhodefexp}
\end{equation}
which contains the scale radius, $r_s$ which is calculated from the virial radius, $R_v(M)$ of the dark matter halo, and the concentration parameter of the HI, $c_{\rm HI} (M,z)$,  as:
\begin{equation}
 r_s = R_v(M,z)/c_{\rm HI} (M,z)
 \end{equation}
where the concentration parameter can be expressed as:
\begin{equation}
 c_{\rm HI}(M,z) =  c_{\rm HI, 0} \left(\frac{M}{10^{11} M_{\odot}} \right)^{-0.109} \frac{4}{(1+z)^{\gamma}}.
\end{equation}
The constant $\rho_0$ in the Eq. (\ref{rhodefexp}) is fixed by normalizing the HI profile within the virial radius $R_v$ to be equal to $M_{\rm HI}$. Hence, the two free parameters in the HI density distribution are $c_{\rm HI, 0}$ and $\gamma$.
{The above formalism is justified by simulations and observations of HI in DLAs, e.g., \citet{barnes2010, barnes2014, maller2004}. This form has been widely used to describe DLA properties, and evidence for its universality also comes from a match to the HI surface density profiles \citep[e.g.,][]{bigiel2012}. Note, however, that all the models above assume the halo mass-dependence of the concentration parameter to be identical to that for dark matter. As we shall see below, the forecasts which are presently possible with intensity mapping experiments do not efficiently constrain the form of the profile, but primarily the parameters in the HI-halo mass relation. Hence, we have chosen not to modify the form of the profile function in the present study.}

Using the above formalism for the HI - halo mass relation and the HI profile, we can compute the power spectrum of the HI intensity fluctuations by defining the Fourier transform of the density profile:
\begin{equation}
u_{\rm HI}(k|M) = \frac{4 \pi}{M_{\rm HI} (M)} \int_0^{R_v} \rho_{\rm HI}(r) \frac{\sin kr}{kr} r^2 \ dr
\end{equation}
where the profile is assumed truncated at the virial radius of the host halo. From this, we can compute the one- and two-halo terms of the HI power spectrum as:
\begin{equation}
P_{\rm HI} (k,z) = P_{\rm 1h, HI} + P_{\rm 2h, HI}
\end{equation}
where
\begin{equation}
P_{\rm 1h, HI} (k,z) = \frac{1}{\bar{\rho}_{\rm HI}^2} \int dM \  n(M) \ M_{\rm HI}^2(M) \ |u_{\rm HI} (k|M)|^2
\label{power1h}
\end{equation}
and
\begin{eqnarray}
&& P_{\rm 2h, HI}  (k,z) =  P_{\rm lin} (k) \nonumber \\
&&\left[\frac{1}{\bar{\rho}_{\rm HI}} \int dM \  n(M) \ M_{\rm HI} (M) \ b (M) \ |u_{\rm HI} (k|M)| \right]^2
\label{power2h}
\end{eqnarray}
In the above expressions, $n(M)$ denotes the dark matter halo mass function [taken to have the Sheth-Tormen \citep{sheth2002} form in the present study], and $b(M,z)$ \citep{scoccimarro2001} is the corresponding halo bias.
From the above expression for the  power spectrum, we can define the the angular power spectrum, denoted by $C_{\ell}$'s \citep[e.g.,][]{battye2012, seehars2016} which is given by:
\begin{eqnarray}
			 &C_\ell(z,z')& = \frac {2} {\pi} \int d\tilde z\, W(\tilde z) D(\tilde z) \int d\tilde z'\, W'(\tilde z')  D(\tilde z') \nonumber \\
			& &\times  \int k^2 dk\, P_{ \rm HI}(k,z, z') j_\ell(kR(\tilde z)) j_\ell(kR(\tilde z')),
	\label{eq:lintheory}
\end{eqnarray}
where the $W, W'$ are the window functions at the redshifts $z$ and $z'$, taken to be uniform across the redshift bin considered, $R(z)$ is the co-moving distance to redshift $z$, and  $D(z)$ is the growth factor for the dark matter perturbations. {{The power spectrum is normalized to the linear theory matter power spectrum at $z = 0$ and the growth factors are chosen such that $D(0) = 1$.}}
 The calculation of the angular power spectrum can be simplified on using the Limber approximation \citep{limber1953} which is a good approximation in the large $\ell$ limit.

\begin{figure}
\includegraphics[scale = 0.6, width = \columnwidth]{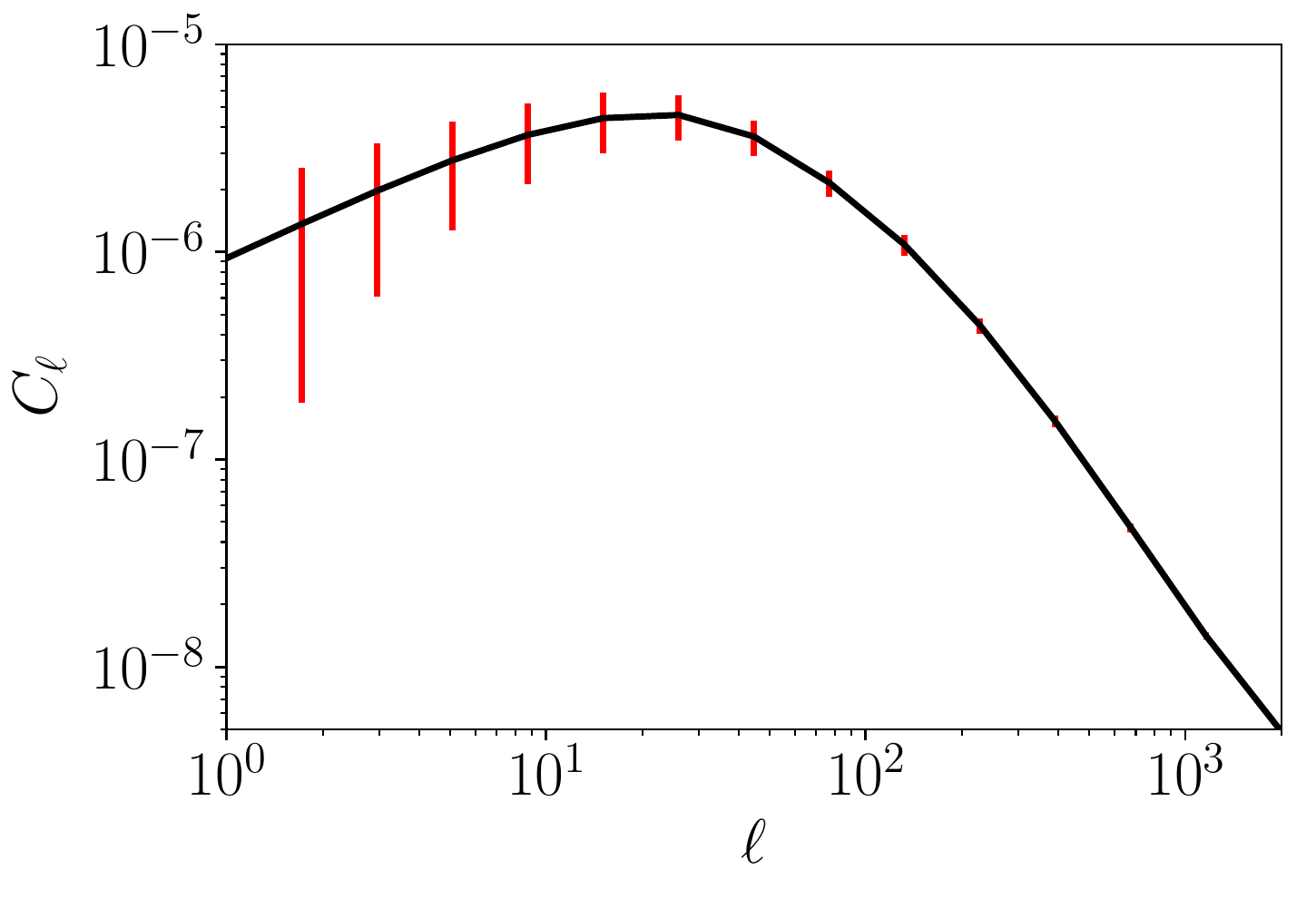}
\caption{Angular power spectrum $C_{\ell}$ from \eq{cllimber} at redshift  0.5, using the fiducial astrophysical and cosmological parameters from Table \ref{table:astrocosmo}. The error bars shown in red represent the standard deviation $\Delta C_l$, calculated following \eq{variance} for the SKA 1 MID configuration. {{Note that the errors on the points below $\ell \sim 10$ are likely to also be affected by the use of the Limber approximation, as mentioned in the main text.}}}
\label{fig:cl}
\end{figure}

 Using the Limber approximation simplifies the $k-$integral in \eq{eq:lintheory} due to the result: 
 \begin{eqnarray}
&&\frac{2}{\pi} \int dk k^2 f(k)  j_\ell(kR(\tilde z)) j_\ell(kR(\tilde z')) = \nonumber \\
\hskip-3in &&  f\left(\frac{\ell + 1/2}{R(\tilde z)} \right) \frac{\delta_D[R(\tilde z') - R(\tilde z)]}{R(\tilde z)^2} \nonumber \\
&\times & \left[1 + \mathcal{O} (\frac{1}{(\ell + 1/2)^2)} \right]
 \end{eqnarray}
where $\delta_D$ denotes the Dirac delta function. The above expression holds for a smooth, not rapidly oscillating $f(k)$ which decreases sufficiently rapidly when $k \to \infty$ \citep[e.g.,][]{bernardeau2012, marozzi2016}. This allows us to rewrite \eq{eq:lintheory} for the large$-\ell$ limit by eliminating the second redshift interval $dz'$:
\begin{equation}
C_{\ell} \simeq \frac{1}{c} \int dz  \frac{W(z)^2 D(z)^2 H(z)}{R(z)^2} P_{\rm HI} [\ell/R(z), z]
\label{cllimber}
\end{equation}
The above approximation is consistent with the findings of \citet{loverde2008}  that the Limber approximation is expected to be accurate to within 1\% above $\ell \sim 10$, for the case of narrow redshift bins such as ours  (see Fig. 1 of \citet{loverde2008}) at redshifts similar to the ones under consideration here. An example angular power spectrum calculated using the above formula is plotted in Fig. \ref{fig:cl}.

Note that the angular power spectra considered in this section are all in real space. We neglect the effects of peculiar velocities which are expected to be unimportant within the noise of these estimates at the scales under consideration.  {{\citet{seehars2016} which uses similar redshift slices as the present study, provides a detailed discussion of the redshift-space effects on the HI angular power spectrum [see Appendix A.3 of \citet{seehars2016}], however, we do not expect these to make a significant difference to the scales of present interest.
}}

As can be seen from the above equation, both  the astrophysical and cosmological parameters enter the expression for the power spectrum. For forecasting the magnitude of the constraints, we adopt a Fisher matrix formalism considering both the mean and variance of the $C_{\ell}$. 
For the comparisons between the various experiments, the following parameters go into the computation of the power spectrum of HI:
\begin{enumerate}
\item  The astrophysical parameters include $v_{c,0}$, $\alpha$, and $\beta$ used in estimating $M_{\rm HI} (M)$, and the normalization $c_{\rm HI, 0}$ and the evolution parameter $\gamma$ used in the HI profile.

\item The cosmological parameters are the Hubble parameter $h$, the baryon density $\Omega_b$, the spectral index $n_s$, the power spectrum normalization parameter $\sigma_8$ and the matter density of the universe, $\Omega_m$.
\end{enumerate}

{{Of the astrophysical parameters mentioned above, it is important to note that only two, viz.  the cutoff and the slope of the HI-halo mass relation, i.e. $v_{c,0}$ and $\beta$ are relevant for forecasting with HI intensity mapping surveys. As can be expected, the parameter $\alpha$ is not constrained by the $C_{\ell}$, this is because it determines the overall normalization and as such cancels in the power spectrum definitions (\eq{power1h} and \eq{power2h}). 
The parameters $c_0$ and $\gamma$ are also found to be poorly constrained by the intensity mapping measurements alone, due to the limited resolution of the experiments under consideration for individual galaxies.}}
Throughout, the cosmology adopted is flat, so that we assume that $\Omega_{\Lambda} = 1 - \Omega_m$. The fiducial values of the parameters are listed in Table \ref{table:astrocosmo}.

 \begin{table}
\begin{center}
    \begin{tabular}{ | c | c | c | c | c | c | c | p{5cm} |}
    \hline
 Astrophysical &  & Cosmological & & \\
 \hline 
 log ($v_{\rm c,0}$/km s$^{-1}$) & 1.56  & $h$ & 0.71  \\
 $\beta$ & -0.58 & $\Omega_m$ & 0.28 \\
 $\alpha$ &  0.09 & $\Omega_b$ & 0.0462 \\
 $c_{\rm HI, 0}$ & 28.65 &  $\sigma_8$ & 0.81\\
 $\gamma$ & 1.45 & $n_s$ & 0.963 \\
 \hline \\
    \end{tabular}
\end{center}
\caption{Fiducial values of astrophysical and cosmological parameters considered. {{The astrophysical parameters come from the best-fitting values of the halo model for neutral hydrogen \citep{hparaa2016} and the cosmological parameters are in good agreement with most available observations, including the latest Planck results \citep{planck}.}}}
 \label{table:astrocosmo}
\end{table}
The Fisher matrix for forecasts on the parameters is computed as follows:
\begin{equation}
F_{ij} = \sum_{\ell} \frac{1}{(\Delta {C_{\ell}})^2}\frac{\partial C_{\ell}}{\partial p_i} \frac{\partial C_{\ell}}{\partial p_j}
\label{fisher}
\end{equation}
where the sum is over the range of $\ell$'s probed, and 
\begin{equation}
(\Delta C_{\ell})^2 = \frac{2 (C_{\ell} + N_{\ell})^2}{(2 \ell + 1)f_{\rm sky}}
\label{variance}
\end{equation}
where the noise term is denoted by $N_{\ell}$ and depends on the particulars of the experiment. If the observing wavelength is denoted by $\lambda_{\rm obs}$, the number of dishes by $N_{\rm dish}$ and the diameter of the dish by $D_{\rm dish}$, the expression for $N_{\ell}$ can be written as \citep[e.g.,][]{battye2012, bull2014}:
\begin{equation}
N_{\ell} = \left(\frac{\sigma_{\rm pix}}{\bar{T}}\right)^2 \left(\frac{\Omega_{\rm pix}}{W_{\ell}}\right)
\label{noise}
\end{equation}
with $W_{\ell} = e^{-\ell^2 \sigma_{\rm beam}^2}$,  $\sigma_{\rm beam} = \theta_{\rm beam}/\sqrt{8 \ln 2}$ and $\theta_{\rm beam} = \lambda_{\rm obs}/(N_{\rm dish} D_{\rm dish})$. The $\bar T(\tilde z)$ is the mean brightness temperature at redshift $z$ defined by:
\begin{equation}
	\bar T(z) \simeq {44} \ {\mu {\rm K}} \left(\frac{\Omega_{\rm HI}(z)h}{2.45\times10^{-4}} \right)\frac{(1+z)^2}{E(z)}
	\label{eq:tbar}
\end{equation}
where $E(z)=H(z)/H_0$ is the normalized Hubble parameter at that redshift {{and $\Omega_{\rm HI} (z)$ is the mean cosmic neutral hydrogen density parameter at redshift $z$}}. The $\Omega_{\rm pix}$ is defined through $\Omega_{\rm pix} = \theta_{\rm beam}^2$, and the $\sigma_{\rm pix}$ is defined by:
\begin{equation}
\sigma_{\rm pix} = \frac{T_{\rm sys}}{\sqrt{t_{\rm pix} \Delta \nu}}
\end{equation}
where $T_{\rm sys}$ is the system temperature, calculated following $T_{\rm sys} = T_{\rm inst} \ +  \ 60 \ {\rm K} \left(\nu/350  \ {\rm MHz} \right)^{-2.5}$ where $T_{\rm inst}$ is the instrument temperature and $\nu$ is the observing frequency. The integration time per beam is $t_{\rm pix}$ (taken to be 1 year for all the surveys considered here\footnote{{This assumption is fairly optimistic given the number of pixels and large sky coverage for some of the surveys under consideration, however the same value is adopted throughout for uniformity. We find by explicit calculation that our main results are  unaffected by the adoption of other, more realistic values of $t_{\rm pix}$.}}) and the $\Delta \nu$ denotes the frequency band channel width, which is connected to the tomographic redshift bin separation $\Delta z$. For the purposes of the noise calculation, we assume $\Omega_{\rm HI}(z)h = 2.45\times10^{-4}$, independent of redshift.
 The fraction of sky probed by the survey, $f_{\rm sky}$ is given by:
\begin{equation}
f_{\rm sky} = \frac{S_{\rm A}}{4 \pi \left(180/\pi \right)^2}
\end{equation}
where the survey area $S_{\rm A}$ is in square degrees.

{{Note that the noise treatment in the preceding discussion is somewhat simplified, using expressions which are formally valid only for single dish receivers (e.g., BINGO, FAST). This is equivalent to replacing the interferometers considered (SKA, CHIME, the planned TianLai) by their effective single-dish configurations. For a compact configuration, it can be shown that the instrument can be treated using an effective single-dish noise expression \citep{seo2010}. Since the present work chiefly focuses on the relative degradation in the constraints due to astrophysical uncertainties, rather than the absolute constraints (which will also be influenced e.g., by other factors such as the cosmological priors adopted), we work with the same noise power expressions in all cases to enable ease of comparison between experiments.}

Other contributors to the total angular power spectrum include the foregrounds, which are likely to be the limiting systematic. Another factor arises from the shot noise of the discrete HI sources; however, it can be shown that for the present context of intensity mapping in the redshift regimes considered here, the shot noise contribution is expected to be negligible \citep{seo2010, seehars2016}.}

Thus, given an experimental configuration specifying the values of $N_{\rm dish}$, $D_{\rm dish}$, the survey area, redshift coverage and instrument temperature, it is possible to compute the Fisher information matrix for a set of cosmological and astrophysical parameters. We now apply this to the various experiments. For completeness, we also  study the comparison between the forecasts derived using the Fisher matrix framework and from a Markov Chain Monte Carlo (MCMC) approach for a few cases in Appendix \ref{sec:fishermcmc}.

We consider six of the forthcoming experiments in the present work:
(i) The Canadian Hydrogen Intensity Mapping Experiment (CHIME)\footnote{https://chime-experiment.ca/}, (ii) BAO In Neutral Gas Observations \citep[BINGO;][]{battye2012}, (iii) TianLai \citep{chen2012}, (iv) the Five hundred metre Aperture Spherical Telescope \citep[FAST;][]{smoot2017}, (v) the Meer-Karoo Array Telescope \citep[MeerKAT;][]{jonas2009} and (vi) the Square Kilometre Array (SKA) Phase I MID \footnote{http://www.ska.ac.za/} (using both bands, B1 + B2). Table \ref{table:parameters} gives the configurations used [for more details on the experiments, see \citet{bull2014}]. More details of the configurations of the BINGO and SKA as regards the noise properties etc. are provided in \citet{olivari2017}.

 \begin{table*}
\begin{center}
    \begin{tabular}{ | c | c | c | c | c | c | c | p{5cm} |}
    \hline
    Configuration & $T_{\rm inst}$ (K) & {{Number of dishes ($N_{\rm dish}$)}} & $D_{\rm dish}$ (m.) & $S_{\rm A}$ (sq. deg.)  & $z_{\rm min}$ & $z_{\rm max}$\\ \hline
    BINGO & 50 & 50 & 25  & 5000 & 0.1 & 0.5\\ 
   CHIME  & 50 & 1280 & 20  & 25000 & 0.8 & 2.5\\ 
   FAST & 20 & 20 & 500  & 2000 & 0.5 & 2.5 \\ 
     TianLai  &   50 & 2048 & 15  & 25000 & 0.5 & 1.55\\
     MeerKAT  &  29 & 64 & 13.5  & 25000 & 0.5 & 1.5\\
     SKA I MID &  28 & 190 & 15  & 25000 & 0.0 & 3.06 \\ \hline
    \end{tabular}
\end{center}
\caption{Various experimental configurations considered in this work.}
 \label{table:parameters}
\end{table*}

\begin{figure*}
\includegraphics[scale = 0.4]{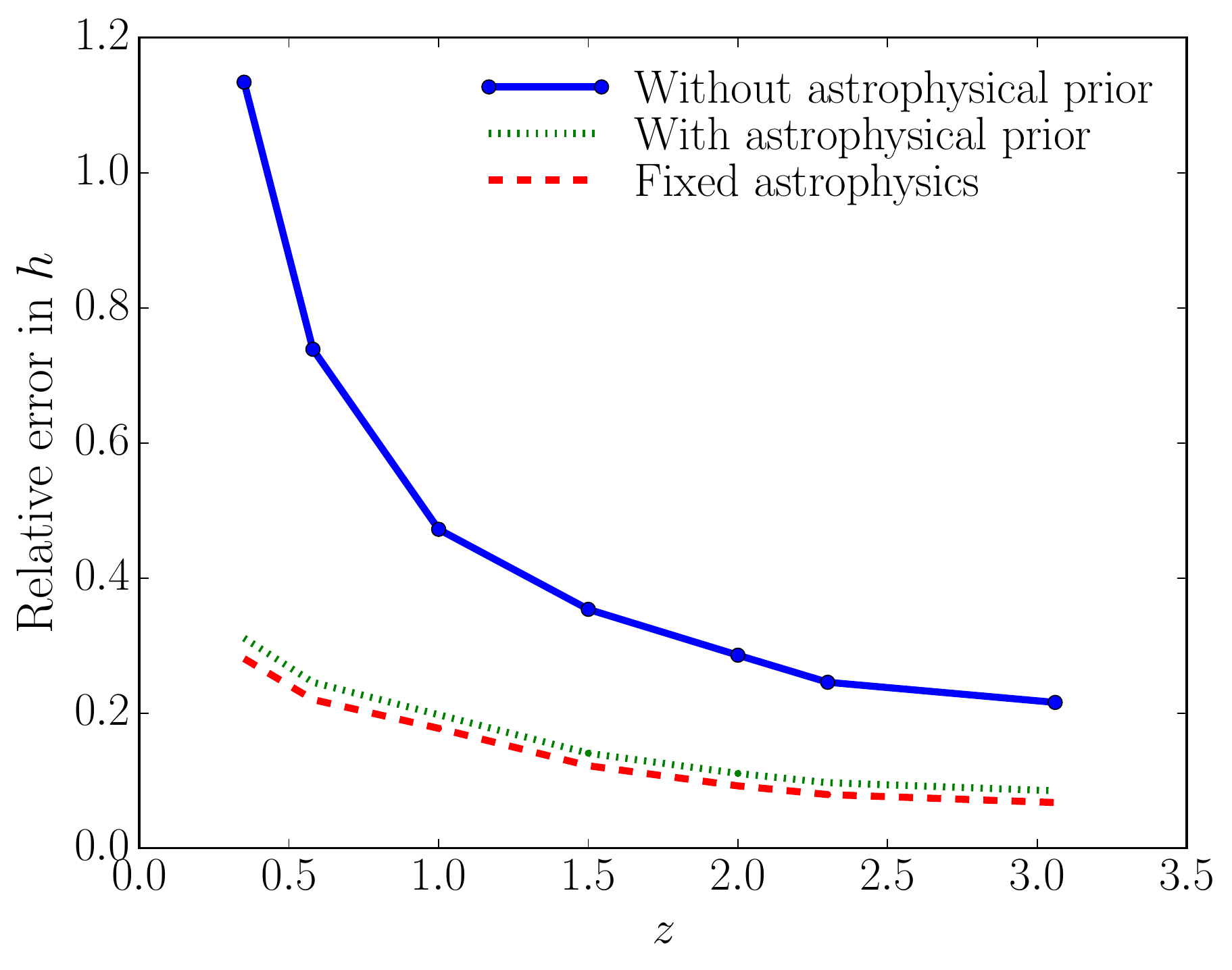} \includegraphics[scale = 0.4]{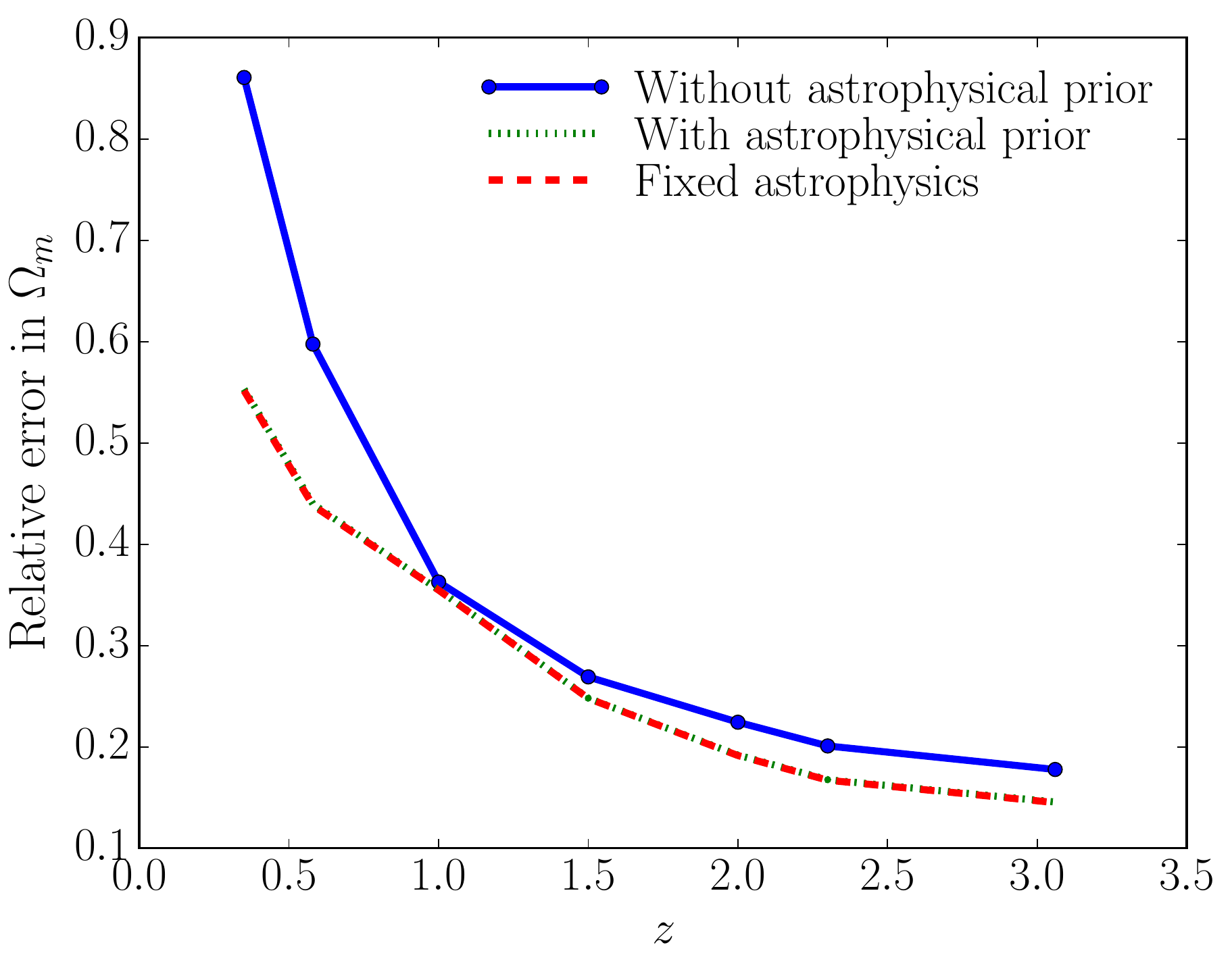} \includegraphics[scale = 0.4]{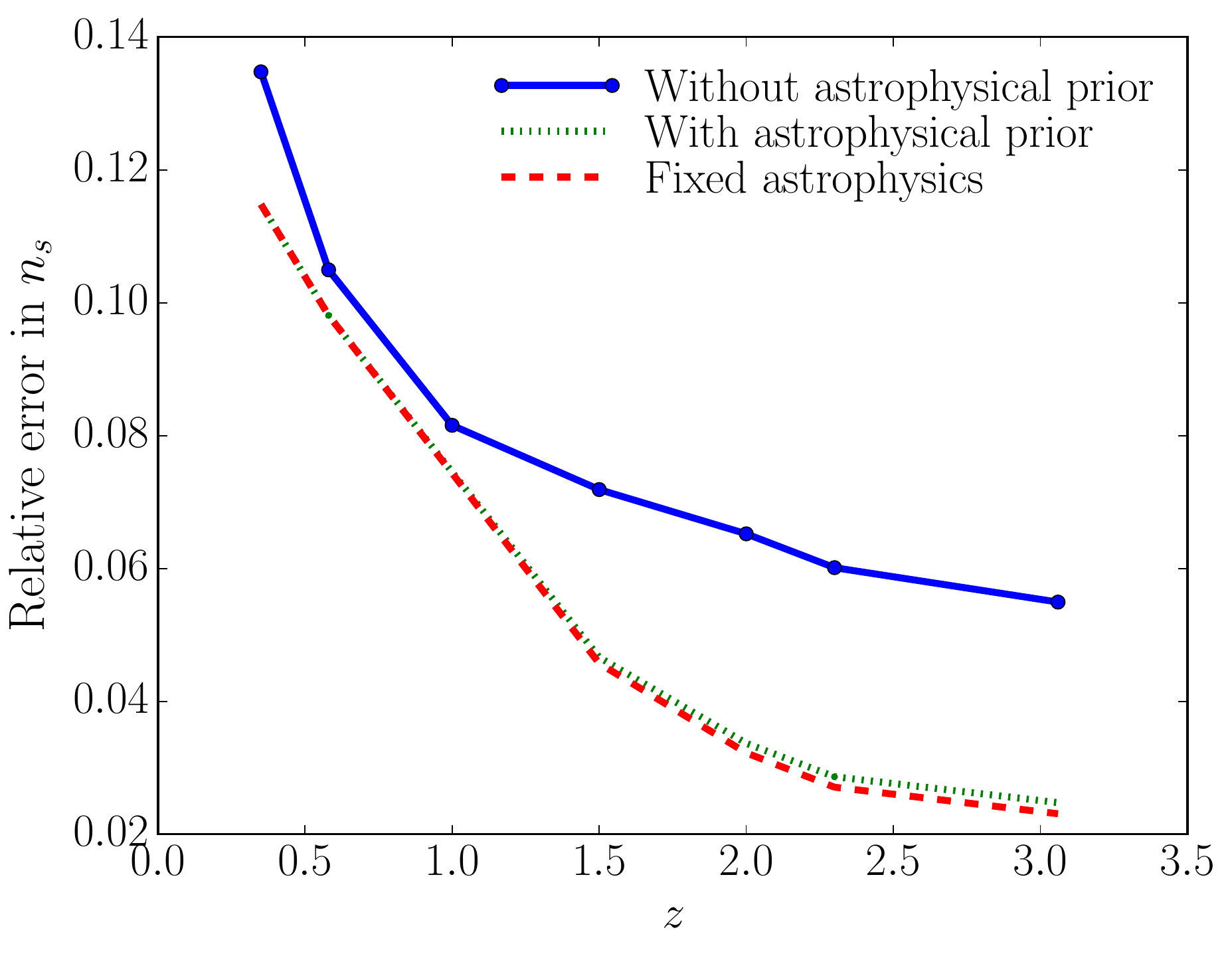} \includegraphics[scale = 0.4]{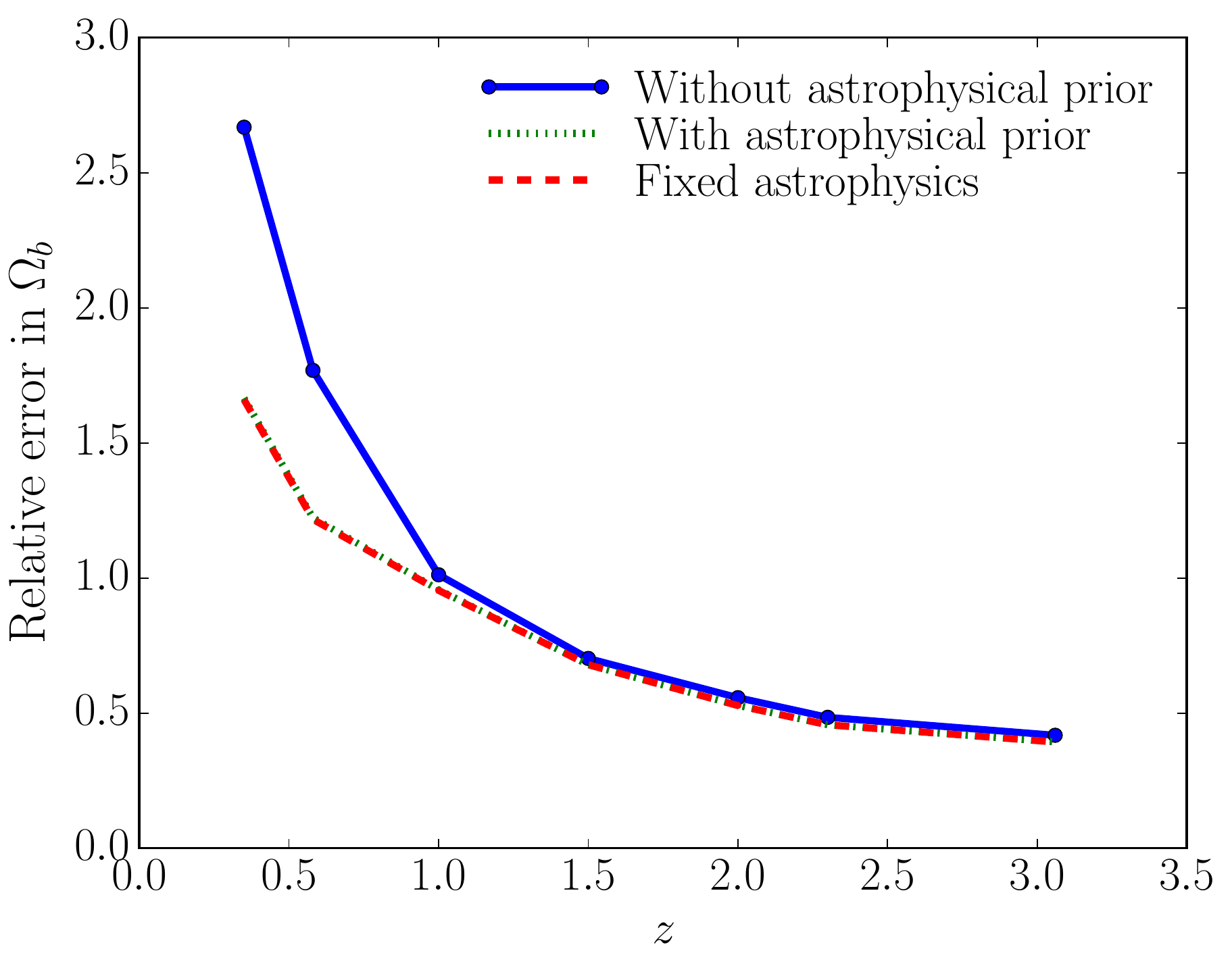} \includegraphics[scale = 0.4]{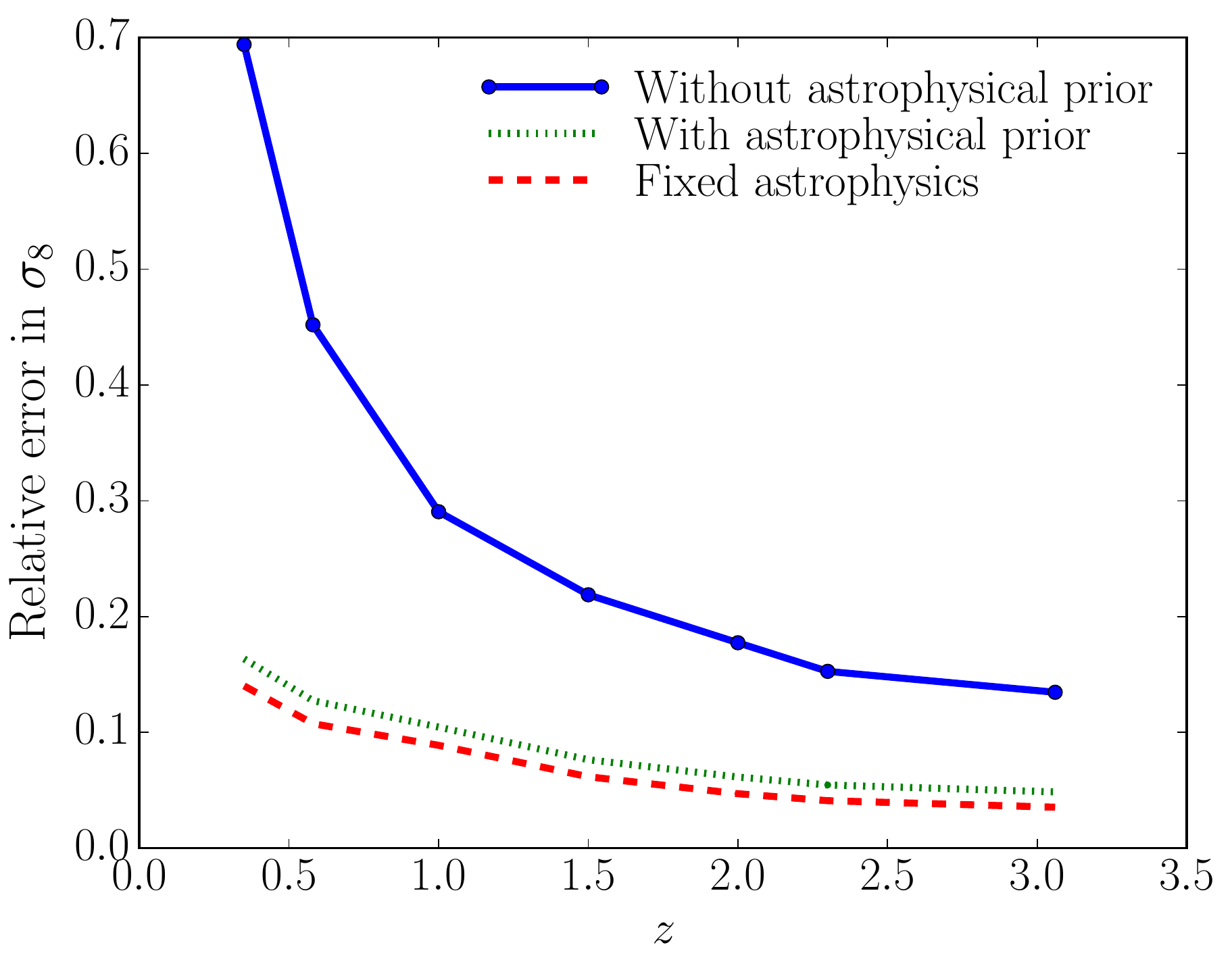} 
\caption{Cosmological forecasts for SKA I MID, (i) with the effects of astrophysical uncertainties (`Without astrophysical prior'), (ii) without the effects of astrophysical uncertainties (`Fixed astrophysics'), and (iii) with the effects of astrophysical uncertainties but also an astrophysical prior added, coming from current knowledge (`With astrophysical prior'). {{Note that in all figures, the results are cumulative as the number of $z-$bins is increased. The scaling with the number of bins is close to, but in some cases significantly steeper than a $1/\sqrt{N(z)}$ form, depending upon the cosmological parameter under consideration.}}}
\label{fig:skanoastro}
\end{figure*}

\begin{figure}
\includegraphics[scale = 0.4, width = \columnwidth]{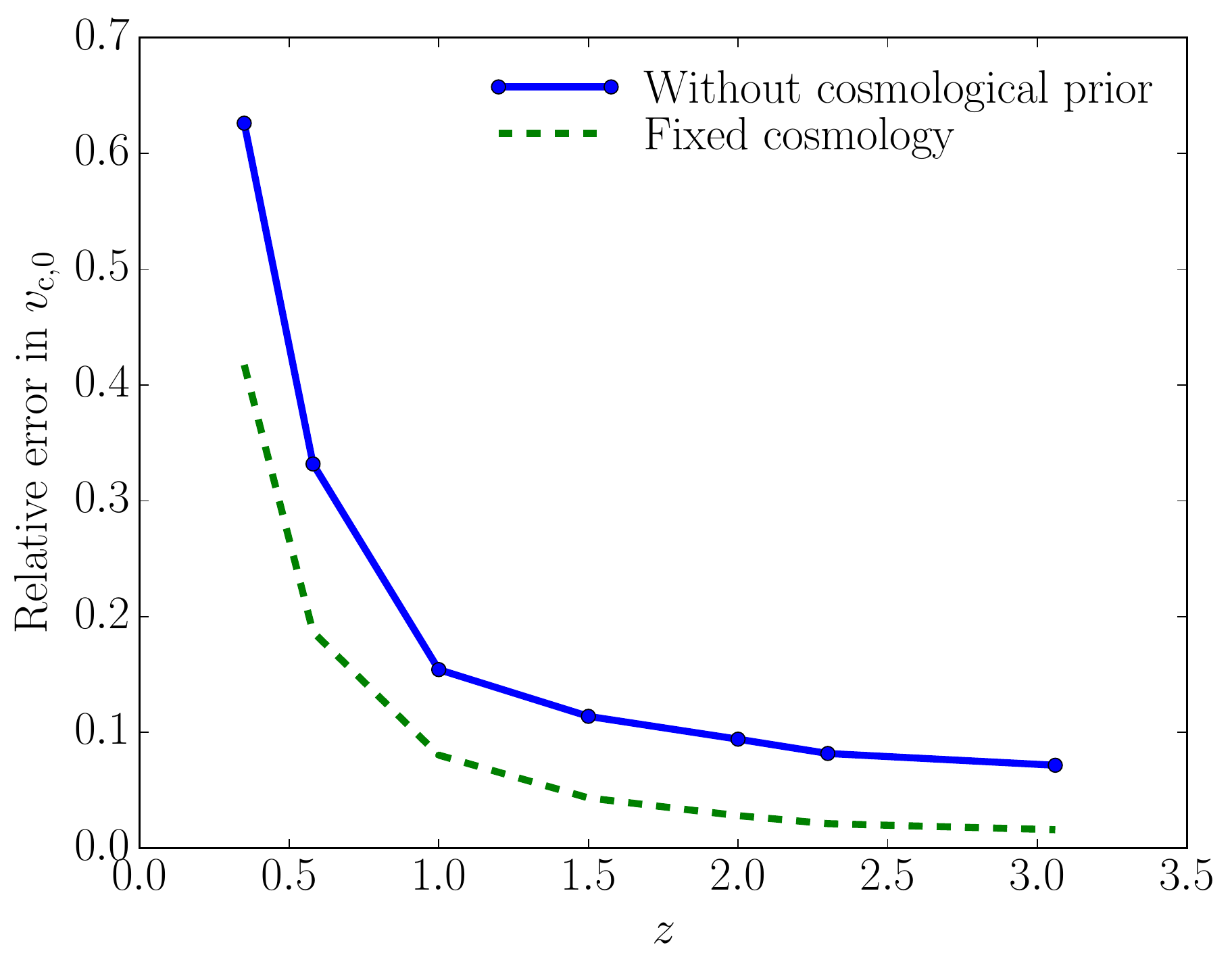} \includegraphics[scale = 0.4, width = \columnwidth]{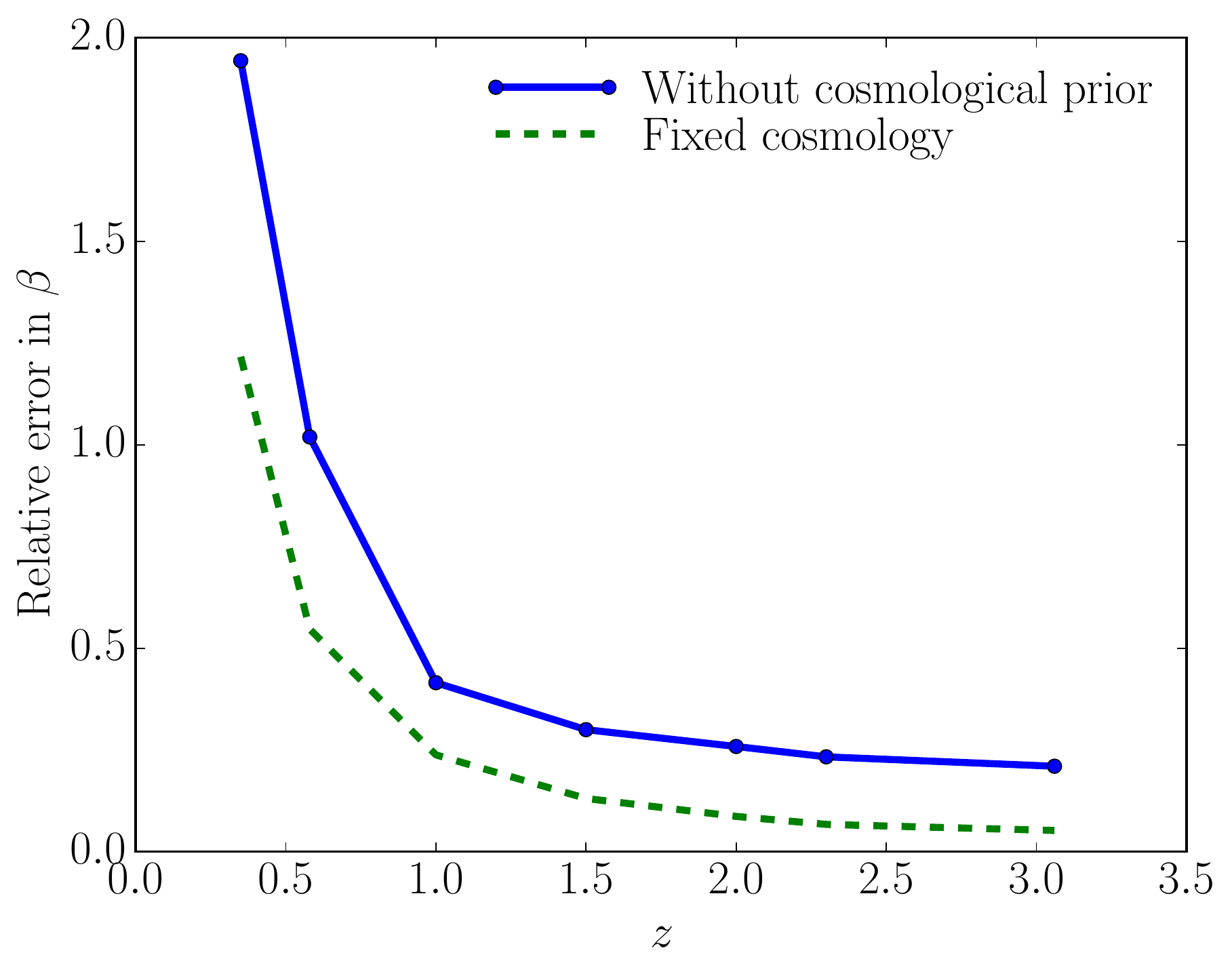}
\caption{Astrophysical forecasts for SKA I MID, (i) with the effects of cosmological uncertainties (`Without cosmological prior'),  and (ii) without the effects of cosmological uncertainties (`Fixed cosmology').{{Note that in all plots, the results are cumulative as the number of $z-$bins is increased. }}}
\label{fig:skanocosmo}
\end{figure}

\section{Fiducial configuration: SKA - I MID: B1 + B2}

In this section, we analyze the effects of the astrophysical uncertainties on the cosmological forecasts in some detail for a particular configuration, namely the SKA I MID, with both Bands 1 and 2\footnote{This leads to the effective redshift range 0 to 3.06.}. 

We compute the noise term as defined by \eq{noise} in the preceding section. We consider equal sized redshift bins of width $\Delta z  = 0.05$ spanning the whole redshift range covered by the experiment, and compute the $C_{\ell}$'s using \eq{cllimber} at the midpoints of each of the redshift bins. Using the values thus obtained, we compute the Fisher forecasts for the parameters  $\Omega_m, n_s, h, \Omega_b, \sigma_8, \beta$ and $v_{c,0}$ from \eq{fisher} for each of the bins. {{We consider the tomographic addition of the bins to derive the cumulative Fisher matrix up to redshift $z$, as given by $F_{ij, {\rm cumul}, z} = \sum_{\Delta z \in z} F_{ij}$  where $F_{ij}$ denotes the Fisher matrix element computed in each redshift bin of width $\Delta z$ included between 0 and $z$. Since the redshift bins are separated by at least 3-5 times the bin width depending on the experiment under consideration, we neglect the effects of cross-correlations between the bins. We use the quantity $F_{ij, {\rm cumul}, z}$ to compute the standard deviations of the various forecasted parameters.}}

The forecasts are shown in Fig. \ref{fig:skanoastro}, for each of the five cosmological parameters, by the blue solid  lines.  In all cases, we see that the tomographic information significantly increases the tightness of the constraints. The saturation occurs after about six or seven redshift bins.

\begin{figure*}
\includegraphics[scale = 0.7]{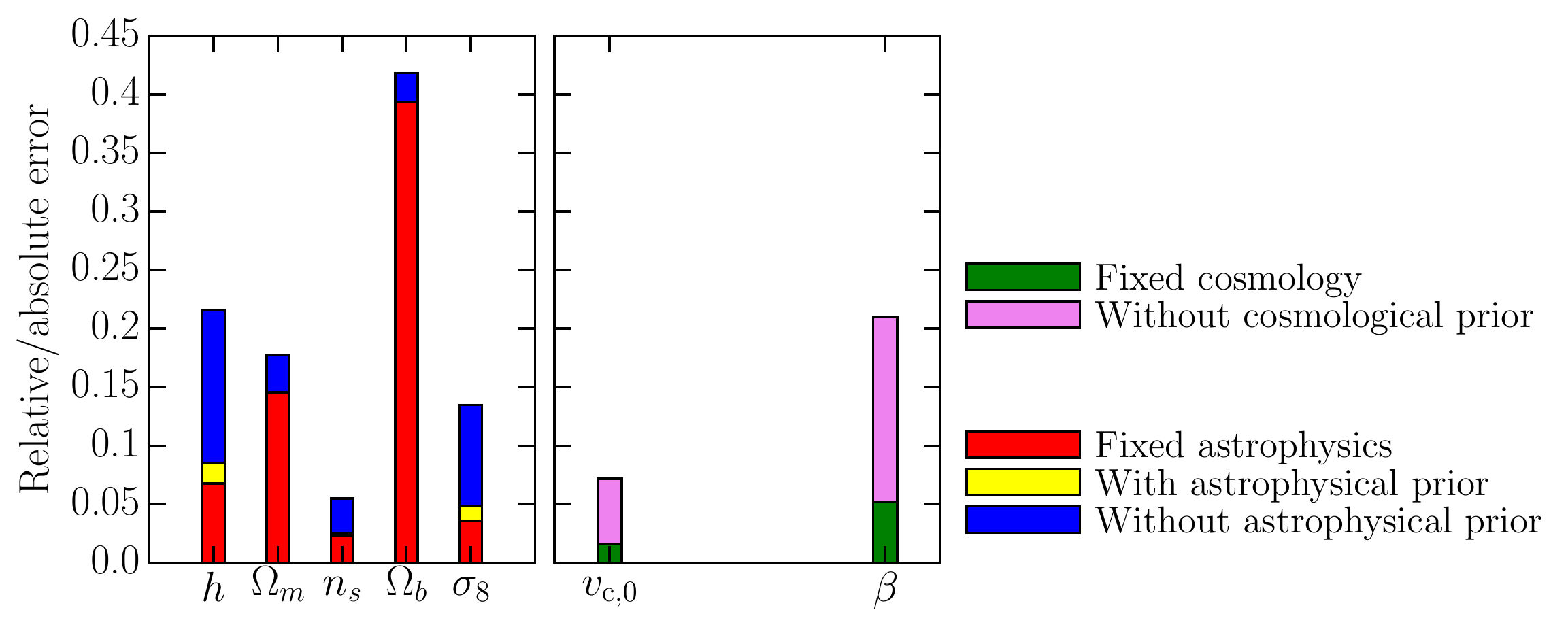} 
\caption{\textit{Left panel:} Asymptotic (best) constraints on the cosmological parameters, (i) without the astrophysical prior, (ii) with fixed astrophysics, and  (iii) with the astrophysical prior coming from the present data, for the case of $\ell_{\rm max} = 1000$. \textit{Right panel:} Astrophysical forecasts for SKA I MID for the case of $\ell_{\rm max} = 1000$, (i) without cosmological priors, and (ii) with fixed cosmology. }
\label{fig:skafid}
\end{figure*}

\subsection{Effect  of  astrophysical priors}

The marginalization over the astrophysical information alone will lead to a degradation in the cosmological parameter values, as compared to the case when the astrophysics is fixed. To better quantify the extent of this degradation (which can be referred to as the `astrophysical systematic'),  we also evaluate the cosmological forecasts without considering the uncertainties in the two astrophysical parameters. The relative errors on each of the cosmological parameters, marginalizing over only the other cosmological parameters, are indicated by the red dashed curves in Fig. \ref{fig:skanoastro}. The figures show that the constraints improve on the addition of the information in different tomographic bins. The best constraints are in the range of 2\% - 50 \%, depending upon the parameter under consideration.

We also consider the effects of the addition of prior knowledge of the astrophysics. To do this, we use the combination of all the data available from the various observations (emission line studies, intensity mapping experiments and Damped Lyman Alpha (DLA) observations) in the post-reionization universe. It was found \citep{hparaa2016} that the currently available data, when combined into a halo model framework, allow fairly stringent constraints on the five astrophysical parameters: $c_{\rm HI, 0} = 28.65 \pm 1.76, \alpha = 0.09 \pm 0.01, \log v_{\rm c,0} = 1.56 \pm 0.04, \beta = -0.58 \pm 0.06, \gamma = 1.45 \pm 0.04$. {{Specifically, we note that \citep{hparaa2016}: (i) there is no evidence for evolution in the HI-halo mass relation apart from the implicit evolution of the virial velocity at fixed halo mass, with redshift, and (ii) the present data disfavor more than 5 parameters to describe the full HI-halo mass and profile including its evolution with redshift. The slope of the HI-halo mass relation, $\beta$, is chiefly sensitive to the HI mass function observations at low redshifts. Constraints on the parameter $c_0$ are mainly driven by the column density distribution of high-redshift DLAs. These are found to be automatically consistent with the surface density profiles observed in low-redshift HI galaxies, e.g. \citet{bigiel2012}. Analyzing a large model space also leads to the above model being picked out as the best-fitting description of the HI data.}}We now use these constraints as priors on the astrophysical parameters.  Adding these priors to the Fisher formalism leads to the results shown with the green dotted curves in Fig. \ref{fig:skanoastro}. These are almost identical to the constraints obtained with the astrophysical parameters fixed to their mean values (red dashed curves, in the same figure). This indicates that the astrophysics is tightly constrained even by the datasets available presently.

As a complementary analysis, we also indicate the constraints on the astrophysical parameters, both  with the marginalization over the cosmological parameters, as well as with the cosmological parameters fixed to their mean values. This is shown in Fig. \ref{fig:skanocosmo}. The figure also shows the constraints for the case of marginalization over only the second astrophysical parameter (denoted as the `Fixed cosmology' case). It can be seen that the constraints in the case of the `Fixed cosmology' improve with the addition of tomographic bins, and saturate as we combine the information from  6 or 7 redshift bins.  
The saturated values of the constraints (the `asymptotic' or `best' constraints) are graphically illustrated in the bar charts of Fig. \ref{fig:skafid}. 

{{It can be seen from Fig. \ref{fig:skanocosmo} that some of the cosmological parameters (e.g., $n_s$, $h$, $\sigma_8$) are much more affected by marginalizing over the astrophysics compared to the others (e.g., $\sigma_8$, $\Omega_b$). For low redshifts, the Fisher marginalized contours on $\Omega_m$ are found to be less sensitive to changes in the astrophysical parameters, while those on $\sigma_8$ are more sensitive to them. Further, the two astrophysical parameters ($v_{c,0}$ and $\beta$) are found to be fairly degenerate with each other.}}

\subsection{Effects of increasing $\ell$-range}

\begin{figure*}
\includegraphics[scale = 0.7]{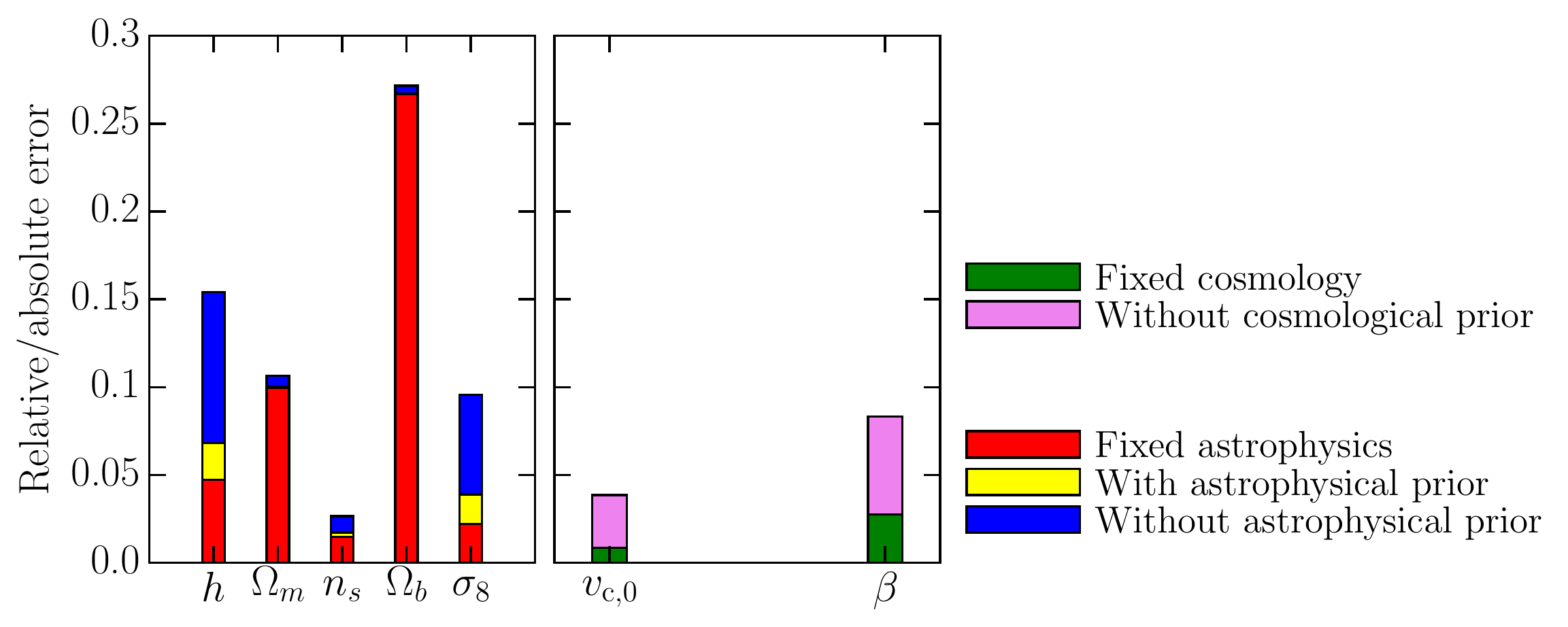} 
\caption{\textit{Left panel:} Asymptotic (best) constraints on the cosmological parameters, (i) without the astrophysical prior, (ii) with fixed astrophysics, and  (iii) with the astrophysical prior coming from the present data, for the case of the extended $\ell$-range, up to $\ell_{\rm max} = 2000$. \textit{Right panel:} Astrophysical forecasts for SKA I MID for the case of $\ell_{\rm max} = 2000$, (i) without cosmological priors, and (ii) with fixed cosmology.}
\label{fig:ska2}
\end{figure*}

 We now explore the  effects of increasing the $\ell$ range for this experiment, to investigate smaller scales (increasing $\ell$ from 1000 to 2000). The asymptotic constraints on the cosmological parameters are shown in the left panel of Fig. \ref{fig:ska2}. All the cosmological constraints are improved by the extension to a larger $\ell$ range. The strong improvement in the parameter $n_s$ is expected, since this parameter is directly connected to the scale $k$.

The best astrophysical constraints with the higher $\ell$-range are shown in the right panel of Fig. \ref{fig:ska2}. Again, the forecasts for both parameters improve on reaching smaller scales.\footnote{As can be expected, the parameter $\alpha$ is not constrained by the $C_{\ell}$, this is because it determines the overall normalization and as such cancels in the power spectrum definitions (\eq{power1h} and \eq{power2h}). 
The parameters $c_0$ and $\gamma$ are also found to be poorly constrained by the intensity mapping measurement alone, however, their constraints also are found to show improvement on extending the $\ell$-range to $\ell_{\rm max} = 2000$.}

\begin{figure*}
 \includegraphics[scale = 0.7]{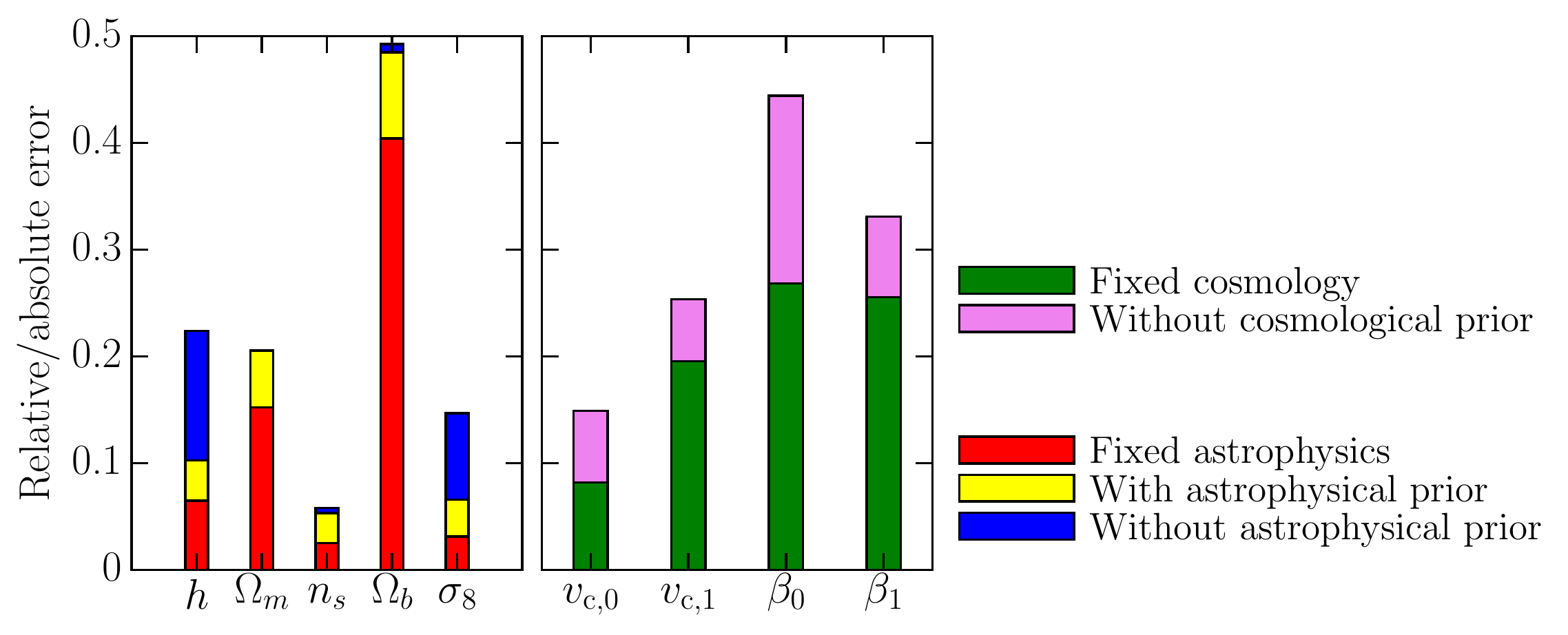} 
\caption{\textit{Left panel:} Cosmological forecasts for SKA I MID with the extended astrophysical parametrization, (i) without the  astrophysical prior, (ii) with fixed astrophysics, and  (iii) with the astrophysical prior coming from the present data. \textit{Right panel:}Astrophysical forecasts for SKA I MID with the extended parametrization, (i) without cosmological priors, and (ii) with fixed cosmology. Relative constraints are shown for the parameters  $v_{c,0}$ and $\beta_0$.  For the new parameters $v_{c,1}$ and $\beta_1$, whose fiducial values are set to zero, the absolute values of the standard deviation obtained by the Fisher analysis are shown. }
\label{fig:skanoastroevol}
\end{figure*}

\subsection{Effects of astrophysical parametrization}
\label{sec:evolution}
Thus far, we have used a parametrization of the HI-halo mass (HIHM) relation which was fitted to the currently available constraints, in the form of \eq{hihm}. However, in the light of the data available from future experiments (such as the SKA I), it may be possible to constrain more parameters of this relation. In this section, we investigate whether (and how) a different (and extended) parametrization of the HIHM affects the results on the forecasts obtained. 

We use, for this purpose, a HIHM relation of the form:
\begin{eqnarray}
M_{\rm HI} (M,z) &=& \alpha f_{H,c} M \left(\frac{M}{10^{11} h^{-1} M_{\odot}}\right)^{\beta(z)} \nonumber \\
&&\times  \exp\left[-\left(\frac{v_{c0}(z)}{v_c(M,z)}\right)^3\right] 
\label{hihmext}
\end{eqnarray}
where 
\begin{equation}
\beta(z) = \beta_0 + \beta_1 \frac{z}{z+1}
\end{equation}
and
\begin{equation}
v_{c0}(z) = v_{c,0} + v_{c,1} \frac{z}{z+1}
\end{equation}
which is a superset of the fiducial HIHM considered in the previous sections. This function reduces to \eq{hihm} when $\beta_1 = v_{c,1} = 0$, with $\beta_0$ reducing to the original $\beta$. Thus, the above form of the HIHM introduces two more free parameters, $v_{c,1}$ and $\beta_1$ into the formalism.

{{For the reasons stated previously in Sec. \ref{sec:fisherforecasts}, it is of interest to consider evolution in the two parameters $\beta$ and $v_{\rm c,0}$ rather than in other parameters such as those related to the profile or the overall normalization of the HI-halo mass relation. Physically, an evolution in $\beta$ represents the possibility of a change in the logarithmic slope of the HI-halo mass relation, and is related to the relative proportion of high-mass halos that serve as HI hosts. The recently reported measurements of the bias of DLAs from cross-correlation analyses with the Lyman-alpha forest \citep{fontribera2012, perezrafols2018} may suggest grounds for such an evolution, however, more data is needed to provide statistical evidence of this \citep{hparaa2016}. Similarly, the parameter $v_{c,0}$ describes the extent of stellar feedback preventing the formation of HI in shallow potential wells \citep{barnes2014, hptrcar2016}. Again, the present data  are consistent with values of $v_{c,0}$ of the order of 30-35 km/s independently of redshift, but higher values would indicate stronger feedback in shallow wells than previously expected, and also shed light on its evolution with redshift.}}

We proceed as in the previous section for the Fisher matrix analysis. The asymptotic (best) relative constraints on the cosmological parameters with this new parametrization are shown in the left panel of Fig. \ref{fig:skanoastroevol}. We also indicate, as before, the cosmological constraints with the astrophysical parameters fixed to their fiducial values (denoted by `Fixed astrophysics'). 
Since the current data do not constrain the values of $v_{c,1}$ and $\beta_1$ \citep{hparaa2016}, we do not consider the effects of astrophysical priors on these two parameters.

We also address the complementary case, i.e. the best constraints on the four astrophysical parameters, in the right panel of Fig. \ref{fig:skanoastroevol}. We also indicate the constraints with the cosmological parameters fixed to their fiducial values (the `Fixed cosmology' case).  Relative constraints are shown  in the cases of the two parameters $v_{c,0}$ and $\beta_0$.  For the new parameters $v_{c,1}$ and $\beta_1$, whose fiducial values are set to zero, we indicate the absolute values of the standard deviation obtained by the Fisher analysis.

Comparison of the left panel of Fig. \ref{fig:skanoastroevol} to that of  Fig. \ref{fig:skafid} reveals that the cosmological constraints are degraded only very weakly by the addition of the two new astrophysical parameters. 
The absolute errors on the quantities $v_{c,1}$ and $\beta_1$ asymptote to values of 0.3. This study indicates, therefore, that the cosmological recovery is  not sensitive to the choice of the astrophysical parametrization used in the analysis. 

\section{Extension to other experiments}

\begin{figure*}
\includegraphics[scale = 0.4]{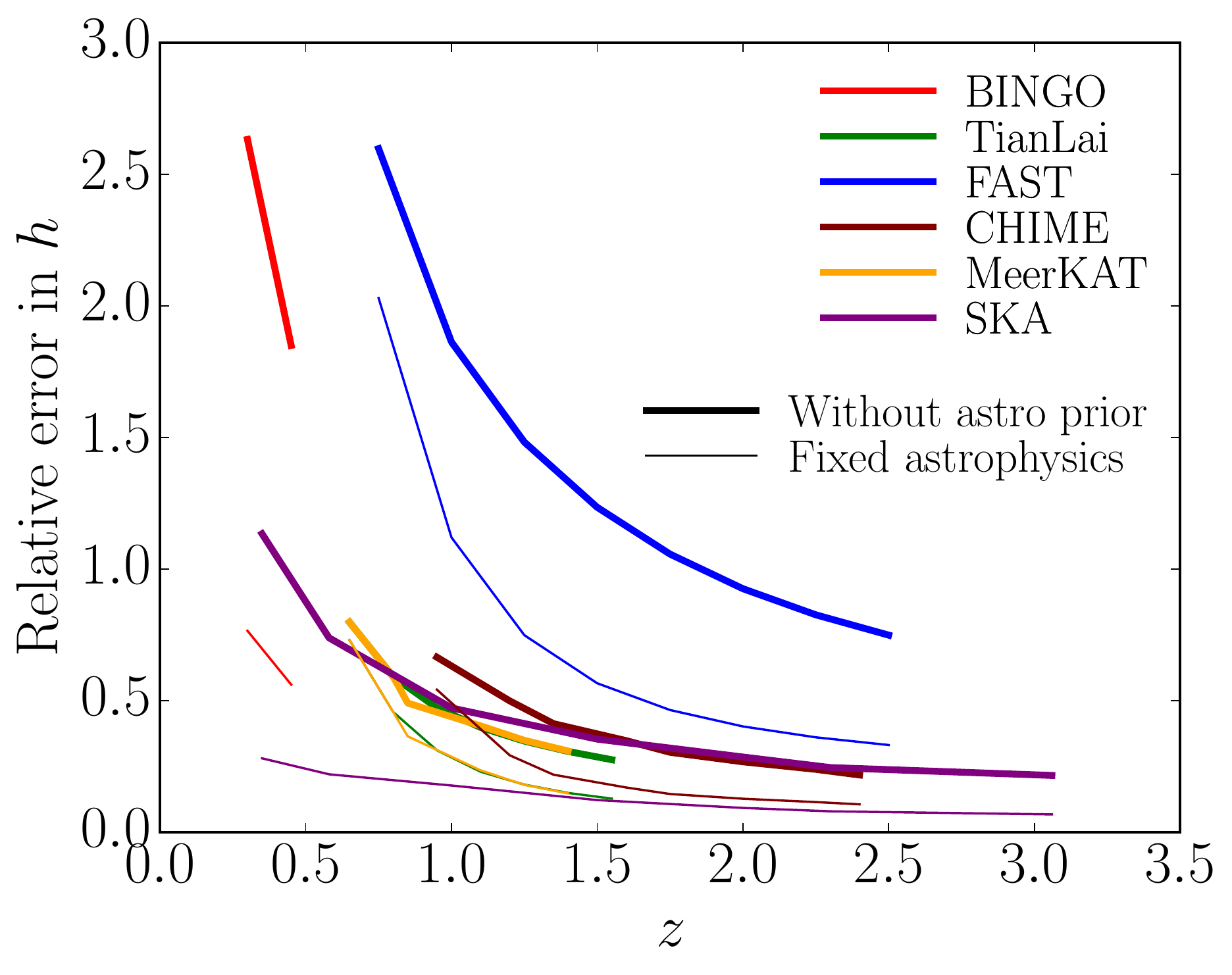} \includegraphics[scale = 0.4]{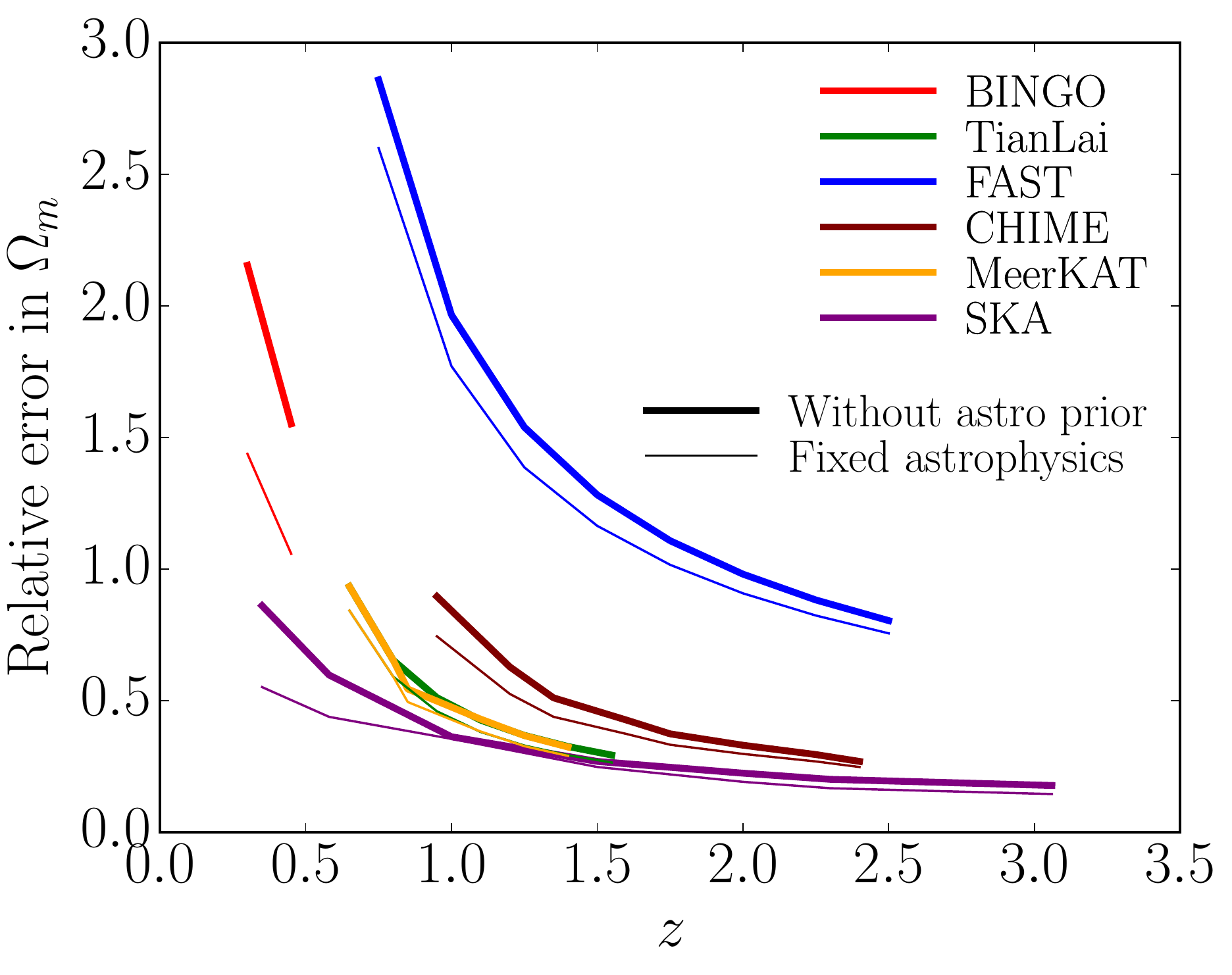}
\includegraphics[scale = 0.4]{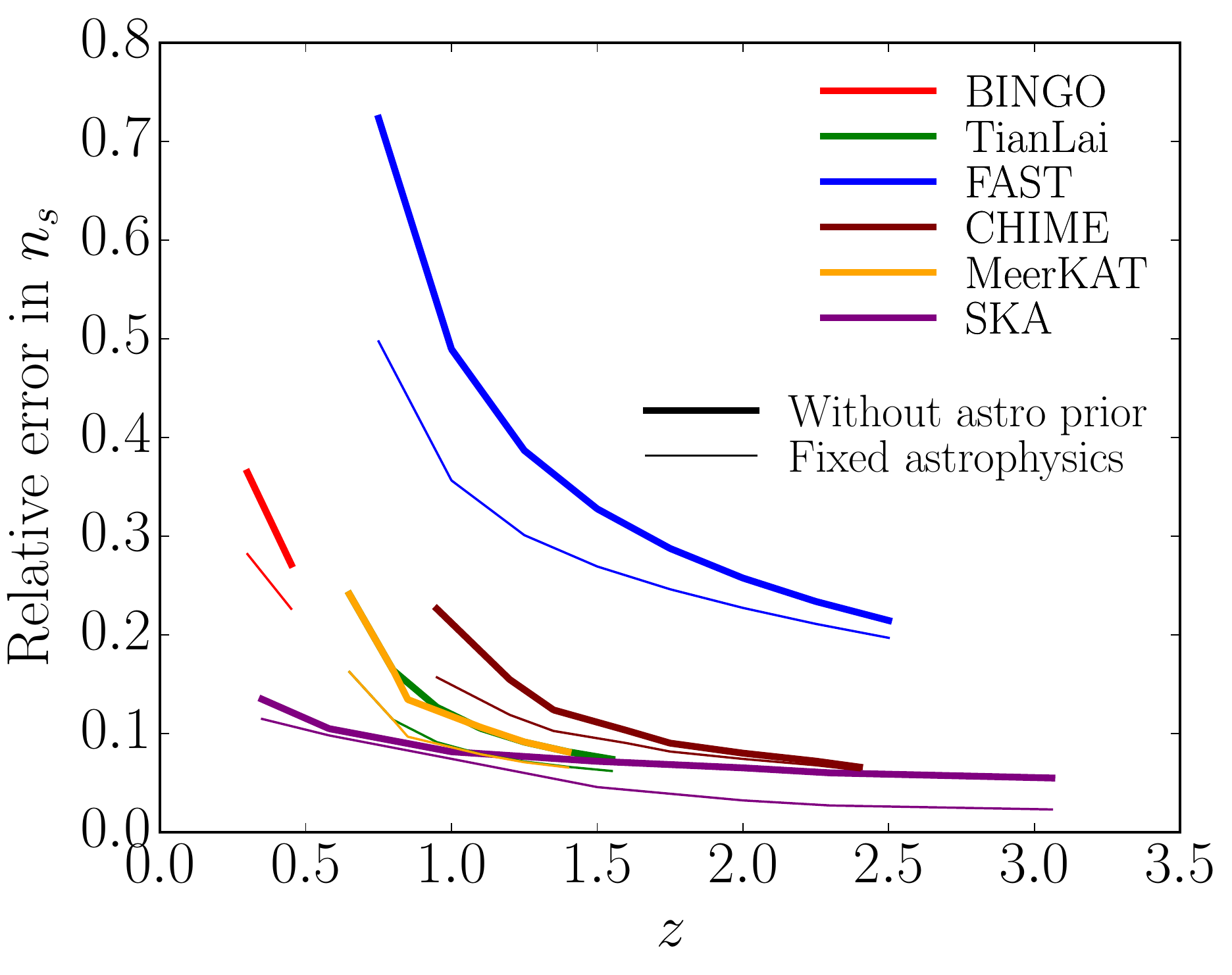} \includegraphics[scale = 0.4]{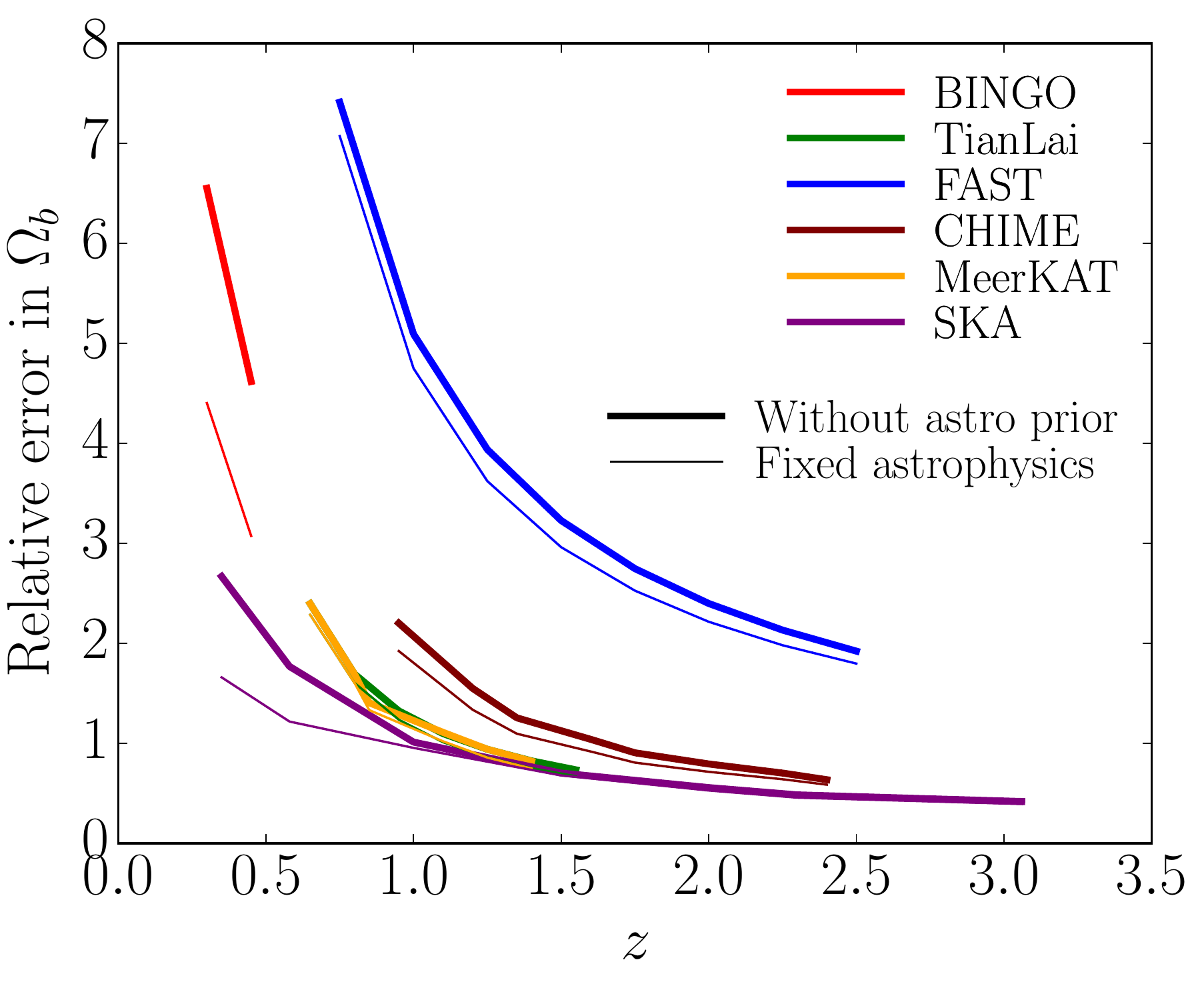}
\includegraphics[scale = 0.4]{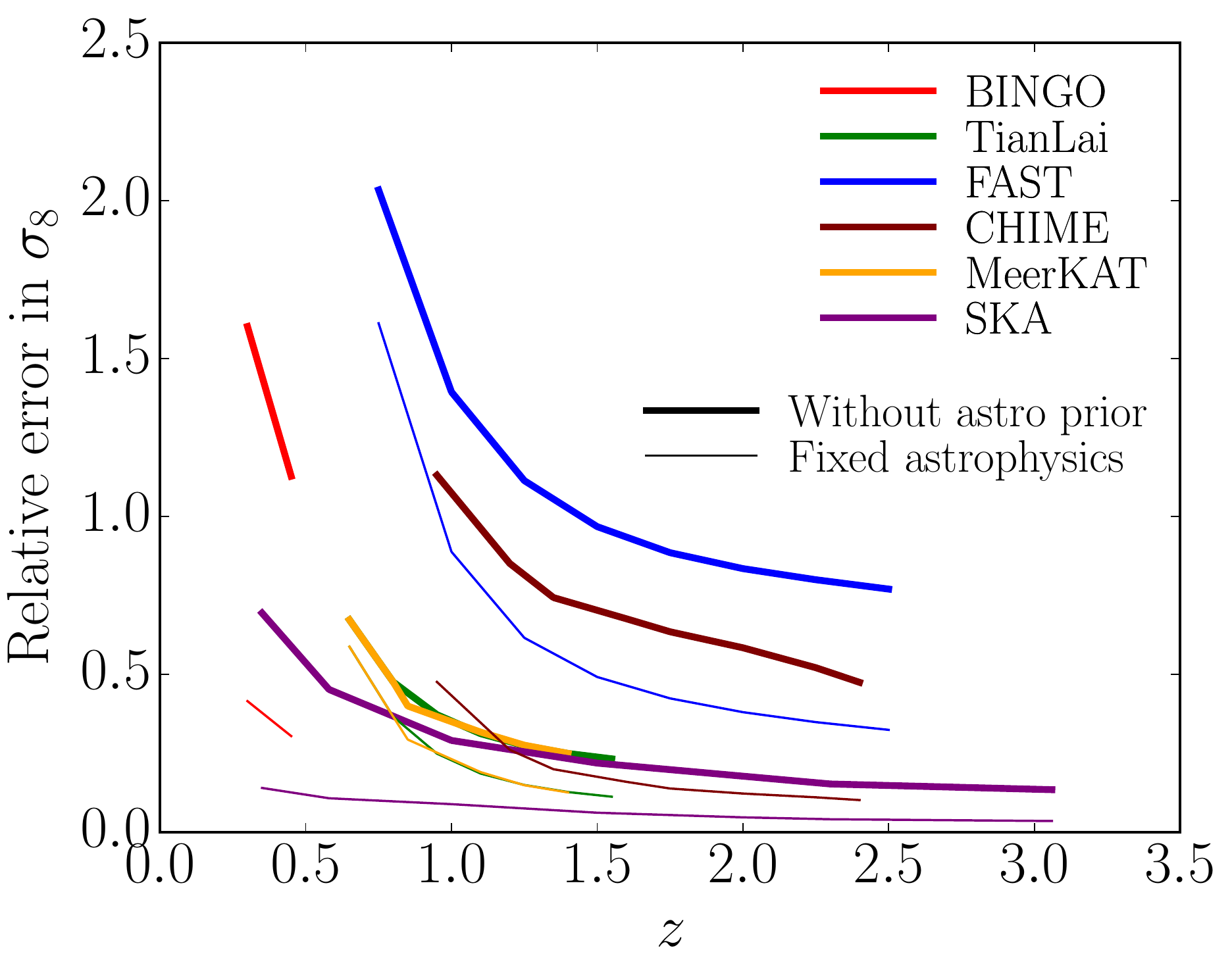}
\caption{Cosmological forecasts for the other experiments -- BINGO (red), TianLai (green), FAST (blue), CHIME (maroon) and  MeerKAT (orange) compared to the fiducial case of SKA I MID (violet). In all cases, the forecasts obtained on marginalizing over all parameters (`Without astro prior') are shown by the thick solid lines. The forecasts with the astrophysical parameters fixed to their mean values (`Fixed astrophysics') are shown by the thin lines for comparison.}
\label{fig:auto1}
\end{figure*}

We now extend the results for the fiducial configuration to the case of the other experiments - BINGO, CHIME, TianLai, MeerKAT and FAST. In each case, we work with a 21 cm autocorrelation intensity mapping survey with the experimental parameters as given in Table \ref{table:parameters}  \citep[see also][]{bull2014}. As in the previous section, we compare the forecasts with and without considering the effects of astrophysical parameters, shown in Fig. \ref{fig:auto1}. The thick lines show the forecasts marginalizing over all the parameters (`Without astrophysical priors'), and the thin lines show the case when the astrophysics is held fixed. We note the following:

\begin{enumerate}
\item As with the fiducial configuration, the cosmological forecasts are affected by the addition of the astrophysical parameters. 

\item The degradation is offset by the increased sensitivity due to the tomographic addition of several redshift bins, saturating, in many cases, with four or five redshift bins.

\item We note the same trend of improvement of the constraints by adding the information from the current knowledge of the astrophysical data (or equivalently, with fixed values of the astrophysical parameters).

{ \item The degree of improvement on adding astrophysical priors is at roughly  the same level for the various experiments (comparing the thin and thick lines of the same color), but the improvement (or conversely, degradation) depends on the cosmological parameter under consideration. This is also seen from the bar charts of Figs. \ref{fig:skafid} and \ref{fig:ska2} in the individual parameters considered. Further investigation e.g. with cross-correlation studies, would shed more light into this inter-relationship and its implications for the astrophysical systematic effects in forecasts.}

\end{enumerate}

\begin{figure}
\includegraphics[scale = 0.6, width = \columnwidth]{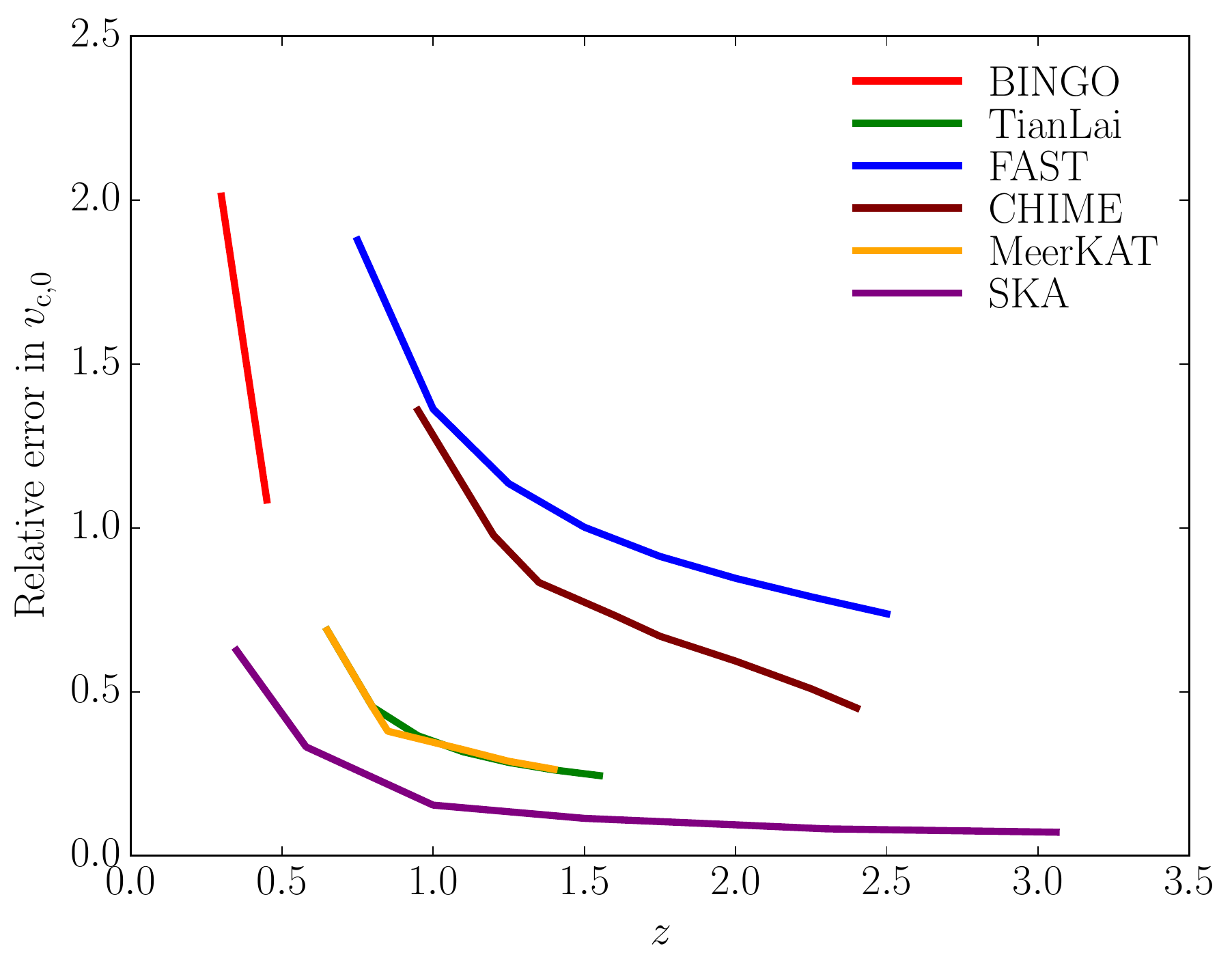}
\includegraphics[scale = 0.6, width = \columnwidth]{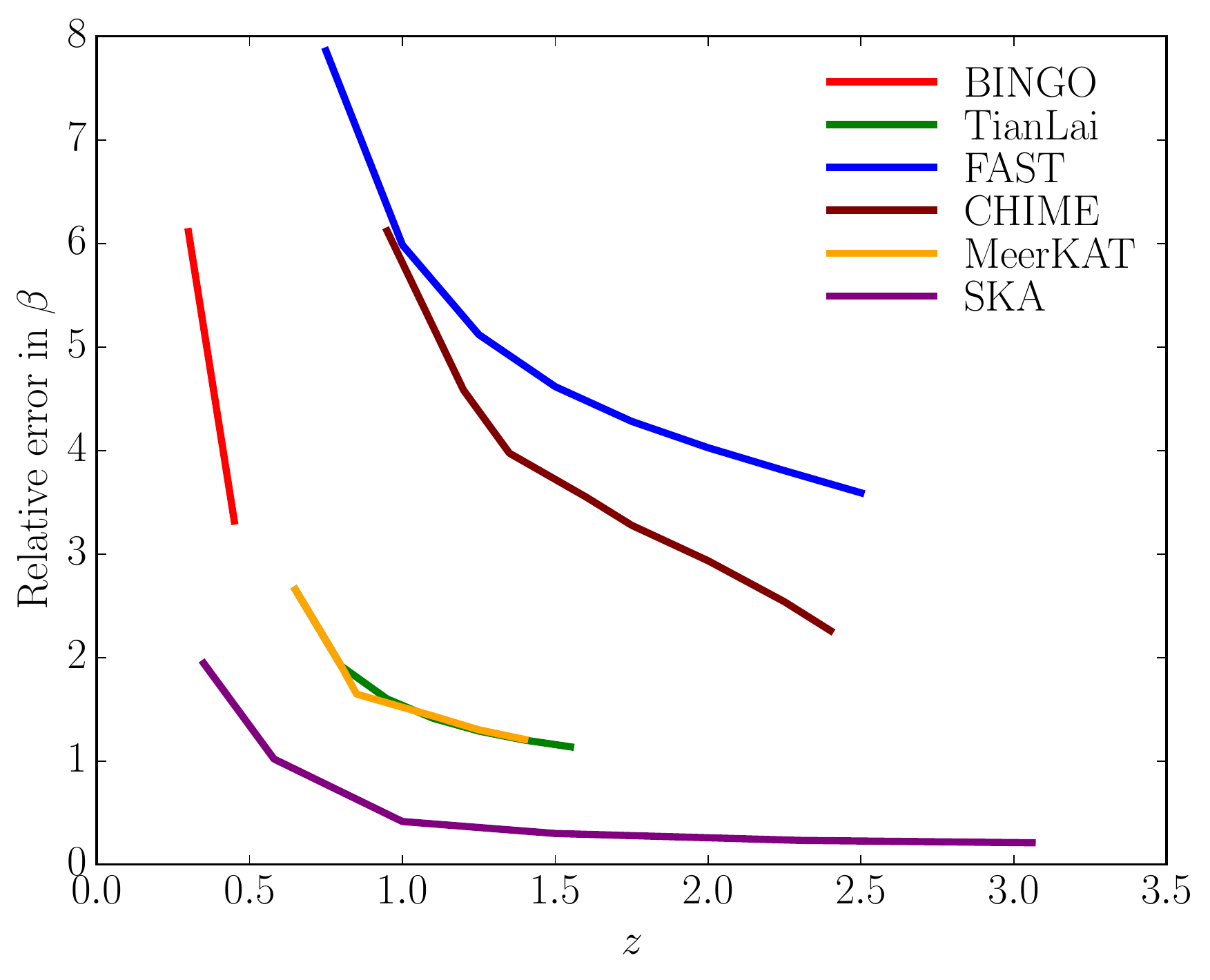}
\caption{Astrophysical forecasts for the other experiments in Fig. \ref{fig:auto1}, marginalizing over the cosmological parameters.}
\label{fig:auto2}
\end{figure}

Fig. \ref{fig:auto2} shows the astrophysical constraints on each of the experimental configurations. Again, the tomographic addition of information from different redshift bins improves the forecasts, just as in the fiducial case considered in the previous sections.

\section{Summary and outlook}

In this paper, we have used the present understanding of the mean and uncertainties in the astrophysical parameters related to neutral hydrogen in the post-reionization universe, to develop forecasts for cosmological and astrophysical parameters with current and future intensity mapping surveys. 

We first considered a particular (`fiducial') experiment, the SKA I MID,  and studied the effect of the astrophysical `systematic' (which needs to be considered in addition to the other systematics caused by instrumental effects and foregrounds). For this experiment and considering scales up to $\ell_{\rm max} = 1000$, we found that marginalizing over the astrophysical parameters (without priors)  broadens the forecasted cosmological constraints. This broadening is by a factor of $2.4 - 2.6$ for the parameters $h$, $n_s$ and $\sigma_8$, and $1.1 - 1.3$ for the parameters $\Omega_b$ and $\Omega_m$. However,  it is, for the large part, alleviated by the addition of prior information coming from our knowledge of astrophysics today.  We studied the robustness of these results to changes in the choice of the astrophysical parametrization considered, and found that an extended HIHM relation did not lead to significant differences in the recovery of the cosmological parameters. Probing smaller scales by increasing $\ell_{\rm max}$ from 1000 to 2000 resulted in a factor of $\sim 1.5 - 1.8$ improvement in the constraints, enabling levels of $4-8\%$ to be reached for the astrophysical parameters $v_{\rm c,0}$ and $\beta$.  We also studied how the constraints improved by increasing the number of tomographic redshift bins, and found saturation, in most cases, with $4 - 5$ redshift bins.

We then compared these results to intensity mapping with other current and future generation facilities, with similar findings. Specifically, for these experiments, the astrophysical uncertainties also cause a broadening in the cosmological constraints, which is, in large part, alleviated by the addition of the prior coming from the current knowledge of the astrophysics. 

We note that the astrophysical uncertainties used for the current data prior are assumed to be dominated by statistical errors (see also \citet{hparaa2016}).  
The effects of systematic errors, the foreground contamination or instrumental effects are not considered in the above forecasts, the primary aim being to explore the inherent broadening in the parameters due to the present state of knowledge of the astrophysics which goes into the HI halo model. 
The cosmological and astrophysical constraints can be improved by the combination of these estimates with the priors from, e.g. CMB experiments {{(which would pin down the errors on e.g., the combinations $\Omega_b h^2$ and $\Omega_m h^2$)}}, as well as cross-correlations with other probes. This  will also help to reduce the systematics from the instrumental effects. Some examples of these have been explored in, e.g., \citet{villaescusa2015, obuljen2017, pourtsidou2017}. We leave the extensions of the present formalism to cross-correlation studies in future work.

\section{Acknowledgements}

HP's research is supported by the Tomalla Foundation. We thank J{\"o}rg Herbel, Raphael Sgier and Christian Monstein for useful discussions. {{We thank the anonymous referee for a detailed and helpful report which improved the content and quality of the presentation.}}

\appendix

\section{Fisher and MCMC comparison}
\label{sec:fishermcmc}

In this appendix, we provide a few examples that illustrate the robustness of the Fisher matrix formalism for forecasting the cosmological parameters in cases where the matrix is well-conditioned (with conditions numbers $\lesssim  100$). Typically, this happens when the low-redshifts ($z < 0.1$) are included in the forecasting, since the constraints are seen to get increasingly stronger at lower redshifts. Here, we focus on the lowest redshift bin, $z   \sim 0.082$, and indicate this comparison for two cases: (i) Joint forecasts on the $\sigma_8$ - $\Omega_m$ plane, and (ii) Joint forecasts on the  $\sigma_8$ - $\Omega_m - \beta$ plane, i.e,  exploring the effect of astrophysical degradation, both using the fiducial SKA I MID (B1 + B2) configuration. We obtain the constraints on the parameters using both the Fisher formalism as described in the main text as well as  a Bayesian Markov Chain Monte Carlo (MCMC) likelihood analysis, and compare the results. 

\begin{figure}
\includegraphics[scale=0.4, width = \columnwidth]{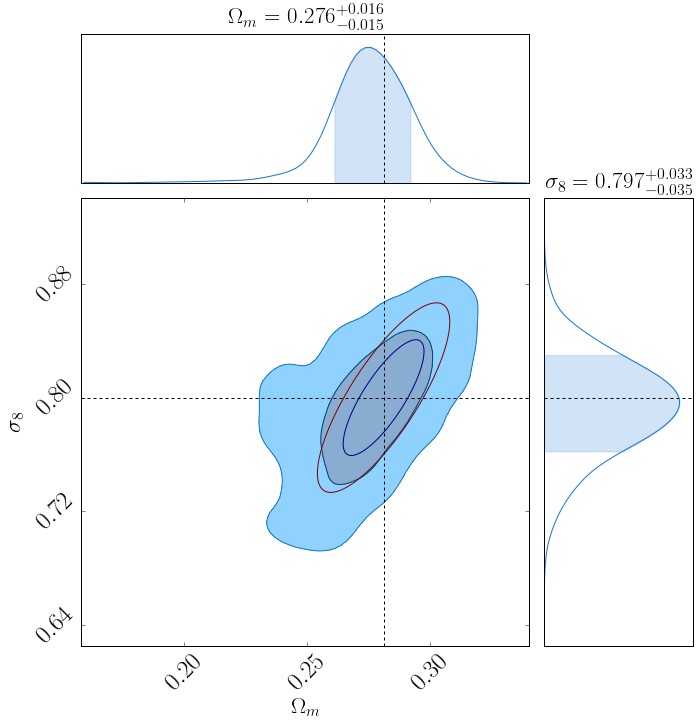}
\includegraphics[scale=0.4, width = \columnwidth]{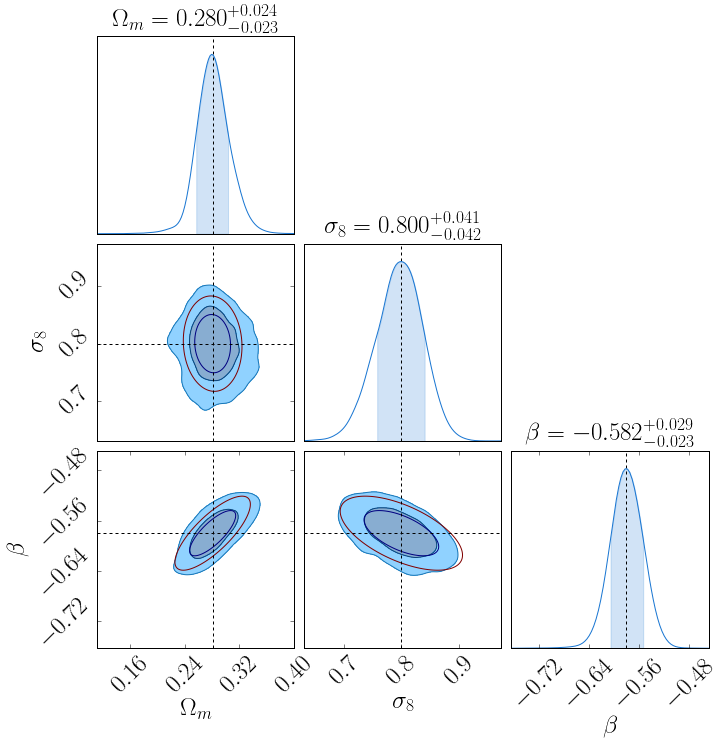}
\caption{Comparison of the constraints obtained with the Fisher and MCMC methods for two illustrative cases at redshift $\sim 0.1$. The top panel shows the constraints on the $\sigma_8 - \Omega_m$ plane, and lower panel shows the case for the $\sigma_8 - \Omega_m - \beta$ parameter space. The dark and light blue shaded regions indicate the 68\% and 95\% levels respectively obtained with the MCMC. The blue and red solid curves indicate the corresponding levels obtained with the Fisher analysis described in the main text.} 
\label{fig:fishermcmc}
\end{figure}

The parameter estimation using MCMC is performed using the likelihood function $\mathcal{L}$ defined through:
\begin{equation}
-2 \ln \mathcal{L} = -\ln \sum_{\ell} \frac{(C_{{\ell}, \rm obs} - C_{{\ell}, \rm calc})^2}{\sigma_{C_{\ell}}^2} 
\label{likelihood}
\end{equation}
In the above expression, the $C_{\ell, {\rm obs}}$ is computed using the best-fitting values of the cosmological and astrophysical parameters. The $\sigma_{C_{\ell}}$ indicates the variance of the angular power spectrum, computed using \eq{variance}. The $C_{\ell, {\rm calc}}$ is the calculated value of the angular power spectrum with the free parameters (i) $\sigma_8$ and $\Omega_m$ and (ii) $\sigma_8$, $\beta$ and $\Omega_m$. The likelihood in \eq{likelihood} is computed using the {\sc{cosmohammer}} package \citep{akeret2013}, and the results are shown in Fig. \ref{fig:fishermcmc}. The  dark and light blue shaded regions indicate the 68\% and 95\% levels respectively obtained with {\sc cosmohammer}. The blue and red solid curves indicate the corresponding constraints obtained with the Fisher analysis. The MCMC and the Fisher forecasts are remarkably similar, thus validating the use of the Fisher formalism in the text for well-conditioned cases.

\bibliographystyle{mnras}
\bibliography{mybib}

\end{document}